\documentclass[australian,showpacs, showkeys,preprint]{revtex4-1}
\usepackage[T1]{fontenc}
\usepackage[latin9]{inputenc}
\usepackage{verbatim}
\usepackage{float}
\usepackage{textcomp}
\usepackage{amsmath}
\usepackage{amssymb}
\usepackage{graphicx}
\usepackage{esint}

\makeatletter

\providecommand{\tabularnewline}{\\}

\makeatother

\usepackage{babel}
\begin{document}

\title{Solution of a generalised Boltzmann's equation for non-equilibrium
charged particle transport via localised and delocalised states}

\author{Peter W. \surname{Stokes}}

\email[Electronic address: ]{peter.stokes@my.jcu.edu.au}

\affiliation{College of Science, Technology and Engineering, James Cook University,
Townsville, QLD 4811, Australia}

\author{Bronson \surname{Philippa}}

\affiliation{College of Science, Technology and Engineering, James Cook University,
Townsville, QLD 4811, Australia}

\author{Daniel Cocks}

\affiliation{College of Science, Technology and Engineering, James Cook University,
Townsville, QLD 4811, Australia}

\author{Ronald D. \surname{White}}

\affiliation{College of Science, Technology and Engineering, James Cook University,
Townsville, QLD 4811, Australia}
\begin{abstract}
We present a general phase-space kinetic model for charged particle
transport through combined localised and delocalised states, capable
of describing scattering collisions, trapping, detrapping and losses.
The model is described by a generalised Boltzmann equation, for which
an analytical solution is found in Fourier-Laplace space. The velocity
of the centre of mass (CM) and the diffusivity about it are determined
analytically, together with the flux transport coefficients. Transient
negative values of the free particle CM transport coefficients can
be observed due to the trapping to, and detrapping from, localised
states. A Chapman-Enskog type perturbative solution technique is applied,
confirming the analytical results and highlighting the emergence of
a density gradient representation in the weak-gradient hydrodynamic
regime. A generalised diffusion equation with a unique global time
operator is shown to arise, reducing to the standard diffusion equation
and a Caputo fractional diffusion equation in the normal and dispersive
limits. A subordination transformation is used to solve the generalised
diffusion equation by mapping from the solution of a corresponding
standard diffusion equation. 
\end{abstract}



\maketitle
\begin{comment}
\tableofcontents{}
\end{comment}

\section{\label{sec:Introduction}Introduction}

Normal transport, as described by the diffusion equation, has a mean
squared displacement that scales linearly with time, $t$. Dispersive
transport, however, is often defined by a mean squared displacement
that scales sublinearly, proportional to $t^{\alpha}$ where $0<\alpha<1$
\citep{metzler2000random}. Physically, this arises due to the presence
of traps causing particles to become immobilised (localised states)
for extended periods of time and resulting in fundamentally slower
transport \citep{scher1975anomalous}. A number of physical systems
have the potential to exhibit dispersive transport. For example, in
organic semiconductors, and other disordered media, trapped states
arise due to local imperfections or variation in the energetic landscape
\citep{scher1975anomalous,sibatov2007fractional}. Electron transport
in certain liquids can be influenced through electrons becoming trapped
in (localised) bubble states (see e.g. \citep{Mauracher2014,Borghesani2002a}),
giving rise to dispersive electronic transport in liquid neon \citep{Sakai1992}.
Similar trapping processes occur for positronium in bubbles (see e.g.
\citep{Stepanov2012c,Dutta2002,Charlton2001}) and positrons annihilation
on induced clusters (see e.g. \citep{Colucci2011}). 

A consequence of dispersive systems, especially those with long-lived
traps, is their dependence on their history. The diffusion equation,
which uses a local time operator, is fundamentally incapable of describing
such memory effects. Mathematically, an adequate model for dispersive
transport requires a global time operator that acts on the entire
history of the system. One successful approach for modelling dispersive
transport is by replacing the the local time derivative in the diffusion
equation with a global fractional time derivative of order $\alpha$
\citep{barkai2001fractional,metzler1999anomalous}. This resulting
fractional diffusion equation describes memory effects while also
satisfying the required sublinear scaling of the mean squared displacement.

However, fractional diffusion equations still share the same spatial
operator as the standard diffusion equation which implies implicitly
an assumption of small spatial gradients. At the same time, the memory
of the initial condition can cause large spatial gradients to persist
for all time. This inconsistency challenges the validity of fractional
diffusion equations. This has been addressed by using phase-space
kinetic models for dispersive transport that make no such assumptions
on the size of spatial gradients \citep{Robson2005,Philippa2014}.
Specifically, these have made use of a Boltzmann equation with a generalisation
of the Bhatnagar-Gross-Krook (BGK) collision operator \citep{Bhatnagar1954},
the standard collision operator in semiconductor physics. In our previous
work \citep{Philippa2014}, trapping and detrapping is considered
equivalent to a BGK collision scattering event occurring after a delay
governed by a trapping time distribution. That study did not consider
scattering as a separate process from trapping, thereby limiting the
model to situations where trapping dominates over scattering. However,
scattering events are key to transport in delocalised states, such
as in the conduction band of a semiconductor. The present study builds
upon previous work by incorporating a genuine scattering model into
a kinetic equation with memory of past trapping events. The new, proposed
model also incorporates loss mechanisms such as charged carrier recombination.

In Section \ref{sec:Model} of this paper, we present a generalised
Boltzmann equation with a BGK collision operator to describe transport
via delocalised states, a delayed BGK operator to model trapping and
detrapping associated with localised (trapped) states, and loss terms
corresponding to free and trapped particle recombination. In Section
\ref{sec:Solution}, we determine an analytical Fourier-Laplace space
solution of this model. This analytical solution is used, among other
things, to determine analytical expressions for phase-space averaged
moments of the generalised Boltzmann equation. Spatial moments provide
transport coefficients describing the the motion of the centre of
mass, while velocity moments are used in conjunction to describe the
particle flux using flux transport coefficients. In Section \ref{sec:Hydrodynamic},
the model is explored in the weak-gradient hydrodynamic regime where
it is shown to coincide with both a standard diffusion equation and
a generalised diffusion equation with history dependence. In Section
\ref{sec:Fractional}, the model is also shown to coincide with a
Caputo fractional diffusion equation in the particular case where
transport is dispersive. In Section \ref{sec:Subordination}, the
solution of the generalised diffusion equation is expressed as a subordination
transformation of the solution of a corresponding standard diffusion
equation. Finally, in Section \ref{sec:Conclusion}, we present conclusions
and possible avenues for future work.

\section{\label{sec:Model}Generalised Boltzmann equation}

\begin{figure}
\includegraphics{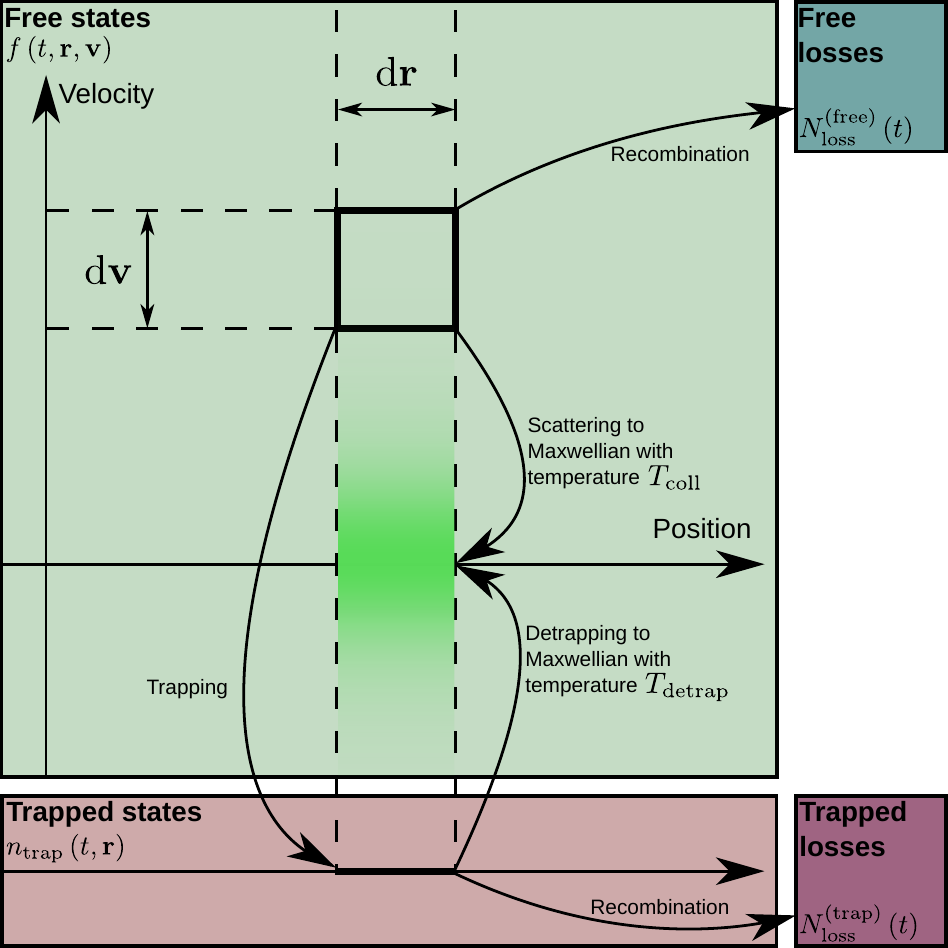}

\centering{}\caption{\label{fig:phaseSpace}Phase-space diagram illustrating the collision,
trapping, detrapping and recombination processes considered in the
model defined by Eqs. (\ref{eq:boltzmannEquation})-(\ref{eq:trapLosses}).}
\end{figure}

We will consider a generalised phase-space kinetic model describing
the transport of free particles undergoing collisions, trapping, detrapping
and recombination as illustrated in Figure \ref{fig:phaseSpace}.
The free particles will be described by the phase-space distribution
function $f\left(t,\mathbf{r},\mathbf{v}\right)$ which satisfies
the Boltzmann equation
\begin{multline}
\left(\frac{\partial}{\partial t}+\mathbf{v}\cdot\frac{\partial}{\partial\mathbf{r}}+\mathbf{a}\cdot\frac{\partial}{\partial\mathbf{v}}\right)f\left(t,\mathbf{r},\mathbf{v}\right)=-\nu_{\mathrm{coll}}\left[f\left(t,\mathbf{r},\mathbf{v}\right)-n\left(t,\mathbf{r}\right)w\left(\alpha_{\mathrm{coll}},v\right)\right]\\
-\nu_{\mathrm{trap}}\left[f\left(t,\mathbf{r},\mathbf{v}\right)-\Phi\left(t\right)\ast n\left(t,\mathbf{r}\right)w\left(\alpha_{\mathrm{detrap}},v\right)\right]-\nu_{\mathrm{loss}}^{\left(\mathrm{free}\right)}f\left(t,\mathbf{r},\mathbf{v}\right),\label{eq:boltzmannEquation}
\end{multline}
where collision, trapping and free particle loss rates are respectively
denoted $\nu_{\mathrm{coll}}$, $\nu_{\mathrm{trap}}$ and $\nu_{\mathrm{loss}}^{\left(\mathrm{free}\right)}$
and the free particle number density is defined $n\left(t,\mathbf{r}\right)\equiv\int\mathrm{d}\mathbf{v}f\left(t,\mathbf{r},\mathbf{v}\right)$.
Collisions are described above by the Bhatnagar-Gross-Krook (BGK)
collision operator \citep{Bhatnagar1954}. Specifically, free particles
are instantaneously scattered to a Maxwellian distribution of velocities
of temperature $T_{\mathrm{coll}}$. The Maxwellian velocity distribution
is defined
\begin{eqnarray}
w\left(\alpha,v\right) & \equiv & \left(\frac{\alpha^{2}}{2\pi}\right)^{\frac{3}{2}}\exp\left(-\frac{\alpha^{2}v^{2}}{2}\right),\\
\alpha^{2} & \equiv & \frac{m}{k_{\mathrm{B}}T},
\end{eqnarray}
where $m$ is the free particle mass, $k_{\mathrm{B}}$ is the Boltzmann
constant and $T$ is the temperature of the scattered particles. Similarly,
trapping and detrapping processes occur as described by a delayed
BGK model \citep{Philippa2014}, according to an effective waiting
time distribution $\Phi(t)$, with trapped particles eventually detrapped
with a Maxwellian velocity distribution of temperature $T_{\mathrm{detrap}}$.
To define this waiting time distribution, consider the simple case
of traps of fixed duration $\tau$. Particles enter traps at the rate
$\nu_{\mathrm{trap}}n\left(t,\mathbf{r}\right)$ and so leave traps
at this same rate $\tau$ units of time in the future. From the present
perspective this rate of detrapping is $\nu_{\mathrm{trap}}n\left(t-\tau,\mathbf{r}\right)$.
More generally, for a distribution of trapping times $\phi\left(t\right)$,
the rate of detrapping is now written as the convolution
\begin{equation}
\nu_{\mathrm{trap}}\phi\left(t\right)\ast n\left(t,\mathbf{r}\right)=\nu_{\mathrm{trap}}\int_{0}^{t}\mathrm{d}\tau\phi\left(\tau\right)n\left(t-\tau,\mathbf{r}\right).
\end{equation}
Here, the quantity $\mathrm{d}P\equiv\phi\left(\tau\right)\mathrm{d}\tau$
can be interpreted as an infinitesimal probability that particles
will remain trapped for duration $\tau$. Note that this expression
does not take into account the possibility that particles may undergo
trap-based losses instead of detrapping. As trapped particles are
being lost exponentially at the rate $\nu_{\mathrm{loss}}^{\left(\mathrm{trap}\right)}$,
the probability of detrapping decays correspondingly, $\mathrm{d}P=\mathrm{e}^{-\nu_{\mathrm{loss}}^{\left(\mathrm{trap}\right)}\tau}\phi\left(\tau\right)\mathrm{d}\tau$.
That is, we have now the effective waiting time distribution
\begin{equation}
\Phi\left(t\right)\equiv\mathrm{e}^{-\nu_{\mathrm{loss}}^{\left(\mathrm{trap}\right)}t}\phi\left(t\right).\label{eq:effectivePhi}
\end{equation}
As the trapped particles are localised in configuration space, we
describe them with the number density $n_{\mathrm{trap}}\left(t,\mathbf{r}\right)$
that satisfies the continuity equation
\begin{equation}
\frac{\partial}{\partial t}n_{\mathrm{trap}}\left(t,\mathbf{r}\right)=\nu_{\mathrm{trap}}\left(1-\Phi\ast\right)n\left(t,\mathbf{r}\right)-\nu_{\mathrm{loss}}^{\left(\mathrm{trap}\right)}n_{\mathrm{trap}}\left(t,\mathbf{r}\right),\label{eq:trappedContinuity}
\end{equation}
where $\nu_{\mathrm{loss}}^{\left(\mathrm{trap}\right)}$ is the loss
rate of trapped particles. Although the loss processes of the free
and trapped particles can occur through various mechanisms (e.g. recombination,
attachment, ...), for simplicity we will refer to all losses as being
due to recombination processes. The number of free and trapped particles
that undergo recombination, $N_{\mathrm{\mathrm{loss}}}^{\left(\mathrm{free}\right)}\left(t\right)$
and $N_{\mathrm{\mathrm{loss}}}^{\left(\mathrm{trap}\right)}\left(t\right)$,
can be counted accordingly
\begin{eqnarray}
\frac{\mathrm{d}}{\mathrm{d}t}N_{\mathrm{\mathrm{loss}}}^{\left(\mathrm{free}\right)}\left(t\right) & = & \nu_{\mathrm{loss}}^{\left(\mathrm{free}\right)}N\left(t\right),\label{eq:freeLosses}\\
\frac{\mathrm{d}}{\mathrm{d}t}N_{\mathrm{\mathrm{loss}}}^{\left(\mathrm{trap}\right)}\left(t\right) & = & \nu_{\mathrm{loss}}^{\left(\mathrm{trap}\right)}N_{\mathrm{trap}}\left(t\right),\label{eq:trapLosses}
\end{eqnarray}
in terms of the number of free and trapped particles, defined by $N\left(t\right)\equiv\int\mathrm{d}\mathbf{r}n\left(t,\mathbf{r}\right)$
and $N_{\mathrm{trap}}\left(t\right)\equiv\int\mathrm{d}\mathbf{r}n_{\mathrm{trap}}\left(t,\mathbf{r}\right)$.

The physical origin of the differences in the functional form of the
waiting time distribution is dependent on the mechanism for trapping.
For example, for amorphous/organic materials, trapping is into existing
trapped states, and the waiting time distribution is calculated from
the density of trapped states (see e.g. \citep{Philippa2014}). For
dense gases/liquids, the trapped states are formed by the electron
itself, and hence the waiting time distribution is dependent on the
scattering, the fluctuation profiles and subsequent fluid bubble evolution
(see e.g. \citep{Cocks2016}).

\section{\label{sec:Solution}Analytical solution of the generalised Boltzmann
equation}

\subsection{Solution in Fourier-Laplace transformed phase-space}

The Boltzmann equation with the BGK collision operator has been solved
analytically in Fourier-Laplace space \citep{Robson1975}. This same
solution technique can be applied to the generalised Boltzmann equation
(\ref{eq:boltzmannEquation}) with the additional processes of trapping,
detrapping and recombination. Applying the Laplace transform in time,
$t\rightarrow p$, and Fourier transform in phase-space, $\left(\mathbf{r},\mathbf{v}\right)\rightarrow\left(\mathbf{k},\mathbf{s}\right)$,
Eq. (\ref{eq:boltzmannEquation}) transforms to
\begin{multline}
\left(\tilde{p}+\tilde{\nu}+\imath\frac{\partial}{\partial\mathbf{s}}\cdot\imath\mathbf{k}+\mathbf{a}\cdot\imath\mathbf{s}\right)f\left(p,\mathbf{k},\mathbf{s}\right)=f\left(t=0,\mathbf{k},\mathbf{s}\right)+\nu_{\mathrm{coll}}n\left(p,\mathbf{k}\right)w\left(\alpha_{\mathrm{coll}},s\right)\\
+\nu_{\mathrm{trap}}\Phi\left(p\right)n\left(p,\mathbf{k}\right)w\left(\alpha_{\mathrm{detrap}},s\right),\label{eq:FLboltzmannEquation}
\end{multline}
where the Fourier-Laplace transformed phase-space distribution function
is
\begin{equation}
f\left(p,\mathbf{k},\mathbf{s}\right)\equiv\int_{0}^{\infty}\mathrm{d}t\int\mathrm{d}\mathbf{r}\int\mathrm{d}\mathbf{v}\mathrm{e}^{-\left(pt+\imath\mathbf{k}\cdot\mathbf{r}+\imath\mathbf{s}\cdot\mathbf{v}\right)}f\left(t,\mathbf{r},\mathbf{v}\right),
\end{equation}
the Fourier-transformed Maxwellian velocity distribution is
\begin{equation}
w\left(\alpha,s\right)\equiv\exp\left(-\frac{s^{2}}{2\alpha^{2}}\right),
\end{equation}
and the following frequencies have been defined
\begin{eqnarray}
\tilde{p} & \equiv & p+\nu_{\mathrm{trap}}\left[1-\Phi\left(p\right)\right]+\nu_{\mathrm{loss}}^{\left(\mathrm{free}\right)},\label{eq:pTilde}\\
\tilde{\nu} & \equiv & \nu_{\mathrm{coll}}+\nu_{\mathrm{trap}}\Phi\left(p\right).\label{eq:nuTilde}
\end{eqnarray}
By writing all vectors in terms of components parallel and perpendicular
to the unit vector $\hat{\mathbf{k}}\equiv\mathbf{k}/k$
\begin{eqnarray}
\mathbf{s}_{\parallel} & = & \left(\mathbf{s}\cdot\hat{\mathbf{k}}\right)\hat{\mathbf{k}},\\
\mathbf{a}_{\parallel} & = & \left(\mathbf{a}\cdot\hat{\mathbf{k}}\right)\hat{\mathbf{k}},\\
\mathbf{s}_{\perp} & = & \mathbf{s}-\mathbf{s}_{\parallel},\\
\mathbf{a}_{\perp} & = & \mathbf{a}-\mathbf{a}_{\parallel},
\end{eqnarray}
the Fourier-Laplace transformed Boltzmann equation (\ref{eq:FLboltzmannEquation})
can be restated as a single first-order differential equation in the
Fourier velocity space variable $s_{\parallel}$
\begin{multline}
\left[\frac{\partial}{\partial s_{\parallel}}-\frac{1}{k}\left(\tilde{p}+\tilde{\nu}+a_{\parallel}\imath s_{\parallel}+\mathbf{a}_{\perp}\cdot\imath\mathbf{s}_{\perp}\right)\right]f\left(p,\mathbf{k},\mathbf{s}\right)=-\frac{f\left(t=0,\mathbf{k},\mathbf{s}\right)}{k}-\frac{\nu_{\mathrm{coll}}}{k}n\left(p,\mathbf{k}\right)w\left(\alpha_{\mathrm{coll}},s\right)\\
-\frac{\nu_{\mathrm{trap}}\Phi\left(p\right)}{k}n\left(p,\mathbf{k}\right)w\left(\alpha_{\mathrm{detrap}},s\right).\label{eq:phaseSpaceDE}
\end{multline}
Finally, solving Eq. (\ref{eq:phaseSpaceDE}) provides the Fourier-Laplace
transformed solution of the generalised Boltzmann equation (\ref{eq:boltzmannEquation}):
\begin{multline}
f\left(p,\mathbf{k},\mathbf{s}\right)=-\frac{1}{k\mu\left(s_{\parallel}\right)}\int_{-\infty}^{s_{\parallel}}\mathrm{d}\sigma\mu\left(\sigma\right)\Big\{ f\left(t=0,\mathbf{k},\sigma,\mathbf{s}_{\perp}\right)+n\left(p,\mathbf{k}\right)\big[\nu_{\mathrm{coll}}w\left(\alpha_{\mathrm{coll}},\sigma,s_{\perp}\right)\\
+\nu_{\mathrm{trap}}\Phi\left(p\right)w\left(\alpha_{\mathrm{detrap}},\sigma,s_{\perp}\right)\big]\Big\},\label{eq:boltzmannSolution}
\end{multline}
written in terms of the integrating factor
\begin{equation}
\mu\left(s_{\parallel}\right)\equiv\exp\left[-\frac{s_{\parallel}}{k}\left(\tilde{p}+\tilde{\nu}+\frac{1}{2}a_{\parallel}\imath s_{\parallel}+\mathbf{a}_{\perp}\cdot\imath\mathbf{s}_{\perp}\right)\right].
\end{equation}
We will use the this analytical expression (\ref{eq:boltzmannSolution})
to evaluate relevant spatial and velocity moments to obtain macroscopic
transport properties.

\subsection{\label{sub:particleNumber}Particle number and the existence of a
steady state}

\begin{figure}
\includegraphics{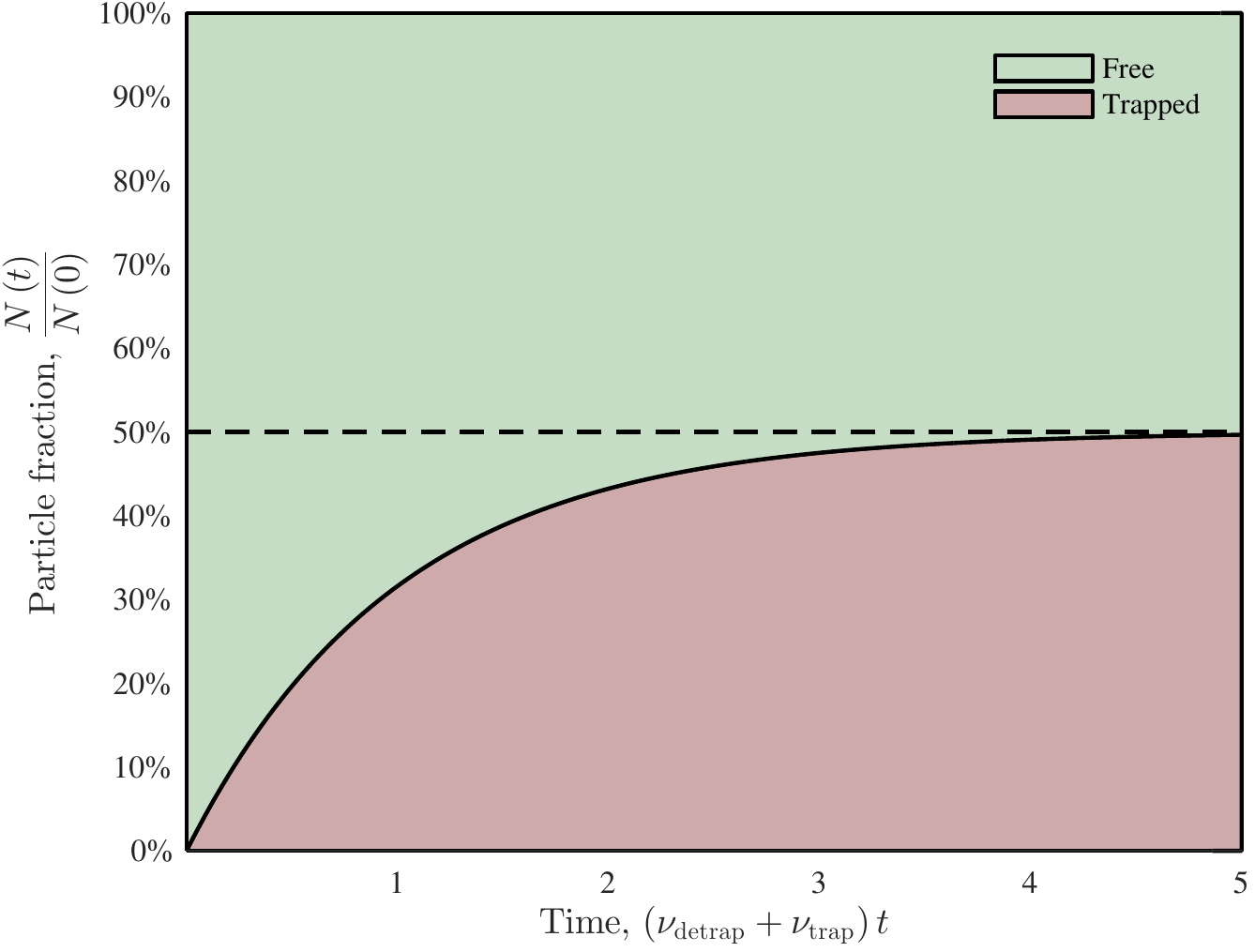}

\caption{\label{fig:steadyState1}Free and trapped particle numbers for the
exponential waiting time distribution $\phi\left(t\right)=\nu_{\mathrm{detrap}}\mathrm{e}^{-\nu_{\mathrm{detrap}}t}$.
As no recombination is present, $\nu_{\mathrm{loss}}^{\left(\mathrm{free}\right)}=\nu_{\mathrm{loss}}^{\left(\mathrm{trap}\right)}=0$,
an equilibrium steady state is reached between the particles as described
by Eqs. (\ref{eq:freeSS}) and (\ref{eq:trapSS}). Here, the detrapping
and trapping rates are set equal, $\nu_{\mathrm{detrap}}=\nu_{\mathrm{trap}}$,
resulting in the same number of free and trapped particles in the
steady state.}
\end{figure}

Integration of the Boltzmann equation (\ref{eq:boltzmannEquation})
throughout all phase-space provides the equation for the free particle
number, $N\left(t\right)$: 
\begin{equation}
\left[\frac{\mathrm{d}}{\mathrm{d}t}+\nu_{\mathrm{trap}}\left(1-\Phi\ast\right)+\nu_{\mathrm{loss}}^{\left(\mathrm{free}\right)}\right]N\left(t\right)=0.\label{eq:particleNumber}
\end{equation}
Similarly, integration over configuration space for the trapped continuity
equation (\ref{eq:trappedContinuity}) provides an equation for the
trapped particle number, $N_{\mathrm{trap}}\left(t\right)$:
\begin{equation}
\left[\frac{\mathrm{d}}{\mathrm{d}t}+\nu_{\mathrm{loss}}^{\left(\mathrm{trap}\right)}\right]N_{\mathrm{trap}}\left(t\right)=\nu_{\mathrm{trap}}\left(1-\Phi\left(t\right)\ast\right)N\left(t\right).
\end{equation}
In conjunction with Eqs. (\ref{eq:freeLosses}) and (\ref{eq:trapLosses})
for the respective number of recombined free and trapped particles,
each particle number can be written explicitly in Laplace space
\begin{eqnarray}
N\left(p\right) & = & \frac{N\left(0\right)}{p+\nu_{\mathrm{trap}}\left[1-\Phi\left(p\right)\right]+\nu_{\mathrm{loss}}^{\left(\mathrm{free}\right)}},\\
N_{\mathrm{trap}}\left(p\right) & = & \frac{\nu_{\mathrm{trap}}\left[1-\Phi\left(p\right)\right]}{p+\nu_{\mathrm{loss}}^{\left(\mathrm{trap}\right)}}N\left(p\right),\\
N_{\mathrm{loss}}^{\left(\mathrm{free}\right)}\left(p\right) & = & \frac{\nu_{\mathrm{loss}}^{\left(\mathrm{free}\right)}}{p}N\left(p\right),\\
N_{\mathrm{loss}}^{\left(\mathrm{trap}\right)}\left(p\right) & = & \frac{\nu_{\mathrm{loss}}^{\left(\mathrm{trap}\right)}}{p}N_{\mathrm{trap}}\left(p\right),
\end{eqnarray}
allowing for steady state values to be determined using the final
value theorem, $\lim_{t\rightarrow\infty}N\left(t\right)=\lim_{p\rightarrow0}pN\left(p\right)$.
Two possible situations arise in the long time limit. In the case
of no recombination, $\nu_{\mathrm{loss}}^{\left(\mathrm{free}\right)}=\nu_{\mathrm{loss}}^{\left(\mathrm{trap}\right)}=0$,
an equilibrium steady state is reached between the free and trapped
particle numbers
\begin{eqnarray}
\lim_{t\rightarrow\infty}\frac{N\left(t\right)}{N\left(0\right)} & = & \frac{\nu_{\mathrm{detrap}}}{\nu_{\mathrm{detrap}}+\nu_{\mathrm{trap}}},\label{eq:freeSS}\\
\lim_{t\rightarrow\infty}\frac{N_{\mathrm{trap}}\left(t\right)}{N\left(0\right)} & = & \frac{\nu_{\mathrm{trap}}}{\nu_{\mathrm{detrap}}+\nu_{\mathrm{trap}}},\label{eq:trapSS}
\end{eqnarray}
where the detrapping rate has been defined
\begin{equation}
\nu_{\mathrm{detrap}}^{-1}\equiv\int_{0}^{\infty}\mathrm{d}t\phi\left(t\right)t.
\end{equation}
Figure \ref{fig:steadyState1} plots the number of free and trapped
particles, $N\left(t\right)$ and $N_{\mathrm{trap}}\left(t\right)$,
and their respective steady state values (\ref{eq:freeSS}) and (\ref{eq:trapSS})
for an exponential waiting time distribution $\phi\left(t\right)=\nu_{\mathrm{detrap}}\mathrm{e}^{-\nu_{\mathrm{detrap}}t}$.

\begin{figure}
\includegraphics{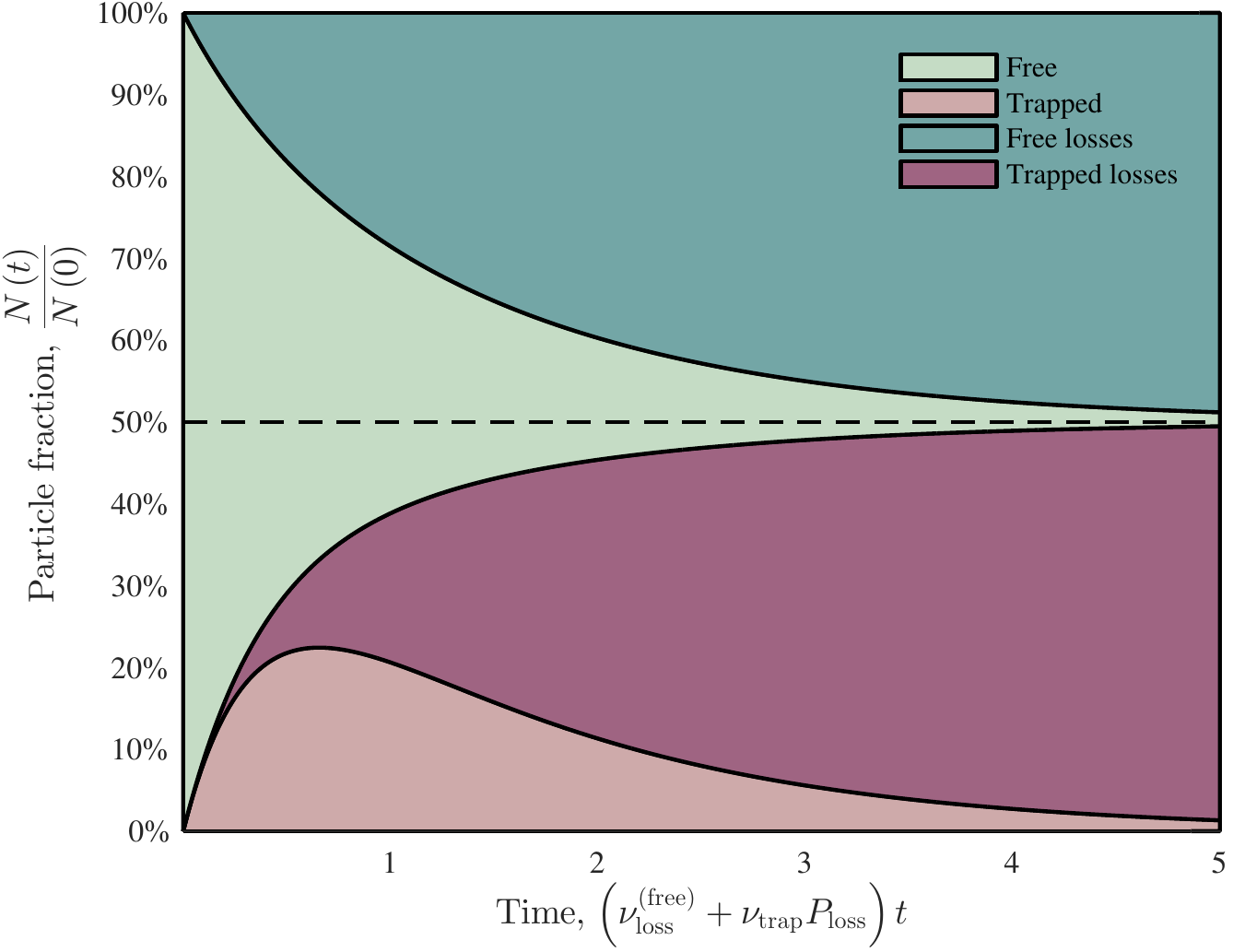}

\caption{\label{fig:steadyState2}Free, trapped and recombined particle numbers
for the exponential waiting time distribution $\phi\left(t\right)=\nu_{\mathrm{detrap}}\mathrm{e}^{-\nu_{\mathrm{detrap}}t}$.
As recombination is present, all free and trapped particles are eventually
lost in the proportions given by Eqs. (\ref{eq:freeLossSS}) and (\ref{eq:trapLossSS}).
Transiently, however, there is an initial increase in the number of
trapped particles. Here, we set equal the free particle recombination
rate and the product of the trapping rate with the trapped particle
recombination probability, $\nu_{\mathrm{loss}}^{\left(\mathrm{free}\right)}=\nu_{\mathrm{trap}}P_{\mathrm{loss}}$.
For this exponential distribution of waiting times this probability
is $P_{\mathrm{loss}}=\nu_{\mathrm{loss}}^{\left(\mathrm{trap}\right)}/\left(\nu_{\mathrm{detrap}}+\nu_{\mathrm{loss}}^{\left(\mathrm{trap}\right)}\right)$.
By making the aforementioned quantities equal, the number of recombined
free and trapped particles also become equal in the long time limit.
In this case, the detrapping and trapping rates are set to $\nu_{\mathrm{detrap}}=\nu_{\mathrm{trap}}=\nu_{\mathrm{loss}}^{\left(\mathrm{free}\right)}+\nu_{\mathrm{trap}}P_{\mathrm{loss}}$,
which consequently specifies the trapped particle recombination rate
$\nu_{\mathrm{loss}}^{\left(\mathrm{trap}\right)}=\nu_{\mathrm{loss}}^{\left(\mathrm{free}\right)}+\nu_{\mathrm{trap}}P_{\mathrm{loss}}$.}
\end{figure}

In the case of \textit{any} recombination, $\nu_{\mathrm{loss}}^{\left(\mathrm{free}\right)}>0$
or $\nu_{\mathrm{loss}}^{\left(\mathrm{trap}\right)}>0$, no free
particle steady state is reached as all free and trapped particles
are eventually lost in the proportions
\begin{eqnarray}
\lim_{t\rightarrow\infty}\frac{N_{\mathrm{loss}}^{\left(\mathrm{free}\right)}\left(t\right)}{N\left(0\right)} & = & \frac{\nu_{\mathrm{loss}}^{\left(\mathrm{free}\right)}}{\nu_{\mathrm{loss}}^{\left(\mathrm{free}\right)}+\nu_{\mathrm{trap}}P_{\mathrm{loss}}},\label{eq:freeLossSS}\\
\lim_{t\rightarrow\infty}\frac{N_{\mathrm{loss}}^{\left(\mathrm{trap}\right)}\left(t\right)}{N\left(0\right)} & = & \frac{\nu_{\mathrm{trap}}P_{\mathrm{loss}}}{\nu_{\mathrm{loss}}^{\left(\mathrm{free}\right)}+\nu_{\mathrm{trap}}P_{\mathrm{loss}}},\label{eq:trapLossSS}
\end{eqnarray}
where the probability that a trapped particle undergoes recombination
instead of detrapping is
\begin{equation}
P_{\mathrm{loss}}\equiv1-\int_{0}^{\infty}\mathrm{d}t\Phi\left(t\right).
\end{equation}
Figure \ref{fig:steadyState2} plots the number of free, trapped and
recombined particles in this case where recombination is present for
the same exponential waiting time distribution used in Figure \ref{fig:steadyState1}.
It can be seen that, although there is an initial increase in the
number of trapped particles, all free and trapped particles are eventually
lost to recombination in the respective proportions (\ref{eq:freeLossSS})
and (\ref{eq:trapLossSS}).

\subsection{Moments and transport coefficients}

In this and later sections we will be predominantly interested in
steady state quantities, independent of the choice of initial condition.
For simplicity, we will assume there are initially $N\left(0\right)$
free particles centred at the origin with a Maxwellian distribution
of velocities of temperature $T_{0}\equiv m/k_{\mathrm{B}}\alpha_{0}^{2}$
\begin{equation}
f\left(t=0,\mathbf{r},\mathbf{v}\right)\equiv N\left(0\right)\delta\left(\mathbf{r}\right)w\left(\alpha_{0},v\right).\label{eq:initialCondition}
\end{equation}
Velocity integration of the generalised Boltzmann equation (\ref{eq:boltzmannEquation})
provides the continuity equation for free particle number density
\begin{equation}
\left[\frac{\partial}{\partial t}+\nu_{\mathrm{trap}}\left(1-\Phi\left(t\right)\ast\right)+\nu_{\mathrm{loss}}^{\left(\mathrm{free}\right)}\right]n\left(t,\mathbf{r}\right)+\frac{\partial}{\partial\mathbf{r}}\cdot\left[n\left(t,\mathbf{r}\right)\left\langle \mathbf{v}\right\rangle \left(t,\mathbf{r}\right)\right]=0.\label{eq:particleBalance}
\end{equation}
This can be solved analytically using the generalised Boltzmann equation
solution (\ref{eq:boltzmannSolution}), yielding 
\begin{equation}
n\left(p,\mathbf{k}\right)=\frac{N\left(0\right)\zeta_{0}\left(\mathbf{k}\right)Z\left[-\left(\tilde{p}+\tilde{\nu}\right)\zeta_{0}\left(\mathbf{k}\right)\right]}{1-\nu_{\mathrm{coll}}\zeta_{\mathrm{coll}}\left(\mathbf{k}\right)Z\left[-\left(\tilde{p}+\tilde{\nu}\right)\zeta_{\mathrm{coll}}\left(\mathbf{k}\right)\right]-\nu_{\mathrm{trap}}\Phi\left(p\right)\zeta_{\mathrm{detrap}}\left(\mathbf{k}\right)Z\left[-\left(\tilde{p}+\tilde{\nu}\right)\zeta_{\mathrm{detrap}}\left(\mathbf{k}\right)\right]},\label{eq:continuitySolution}
\end{equation}
where the plasma dispersion function, $Z\left(\xi\right)$, is defined
\citep{fried1961plasma} 
\begin{equation}
Z\left(\xi\right)\equiv\frac{1}{\sqrt{\pi}}\int_{-\infty}^{\infty}\mathrm{d}x\frac{\mathrm{e}^{-x^{2}}}{x-\xi},
\end{equation}
and each Maxwellian yields a term of the form
\begin{equation}
\zeta\left(\mathbf{k}\right)\equiv\left[2\imath\mathbf{k}\cdot\left(\frac{\imath\mathbf{k}}{\alpha^{2}}-\mathbf{a}\right)\right]^{-\frac{1}{2}}.
\end{equation}
From this analytical solution, phase-space averaged moments of the
generalised Boltzmann equation can be found exactly for all times.
For example, we have the spatial moments 
\begin{eqnarray}
\mathcal{L}\left\{ \frac{N\left(t\right)}{N\left(0\right)}\left\langle \mathbf{r}\right\rangle \left(t\right)\right\}  & = & \frac{\mathbf{a}}{\tilde{p}^{2}\left(\tilde{p}+\tilde{\nu}\right)},\\
\mathcal{L}\left\{ \frac{N\left(t\right)}{N\left(0\right)}\left\langle \mathbf{r}\mathbf{r}\right\rangle \left(t\right)\right\}  & = & \frac{2\mathbf{I}}{\tilde{p}^{2}\left(\tilde{p}+\tilde{\nu}\right)^{2}}\left[\frac{\tilde{p}}{\alpha_{0}^{2}}+\frac{\nu_{\mathrm{coll}}}{\alpha_{\mathrm{coll}}^{2}}+\frac{\nu_{\mathrm{trap}}\Phi\left(p\right)}{\alpha_{\mathrm{detrap}}^{2}}\right]+\frac{2\mathbf{a}\mathbf{a}}{\tilde{p}^{2}\left(\tilde{p}+\tilde{\nu}\right)^{2}}\left(\frac{1}{\tilde{p}}+\frac{2}{\tilde{p}+\tilde{\nu}}\right),\nonumber 
\end{eqnarray}
where the Laplace transform operator has been denoted explicitly here
as $\mathcal{L}$. From these moments, the motion of the centre of
mass (CM) can be described. The CM velocity is defined as the time
rate of change of its position
\begin{equation}
\mathbf{W}_{\mathrm{CM}}\left(t\right)\equiv\frac{\mathrm{d}}{\mathrm{d}t}\left\langle \mathbf{r}\right\rangle \left(t\right),\label{eq:CMdrift}
\end{equation}
while the CM diffusivity is defined as being proportional to the rate
of change of particle dispersion about it
\begin{equation}
\mathbf{D}_{\mathrm{CM}}\left(t\right)\equiv\frac{1}{2}\frac{\mathrm{d}}{\mathrm{d}t}\left[\left\langle \mathbf{r}\mathbf{r}\right\rangle \left(t\right)-\left\langle \mathbf{r}\right\rangle \left(t\right)\left\langle \mathbf{r}\right\rangle \left(t\right)\right].\label{eq:CMdiffusion}
\end{equation}
CM transport coefficients can be defined for the free, trapped and
total particles. Although trapped particles are localised in space
their CM still moves due to repeated detrapping and trapping.

\begin{figure}
\begin{centering}
\includegraphics{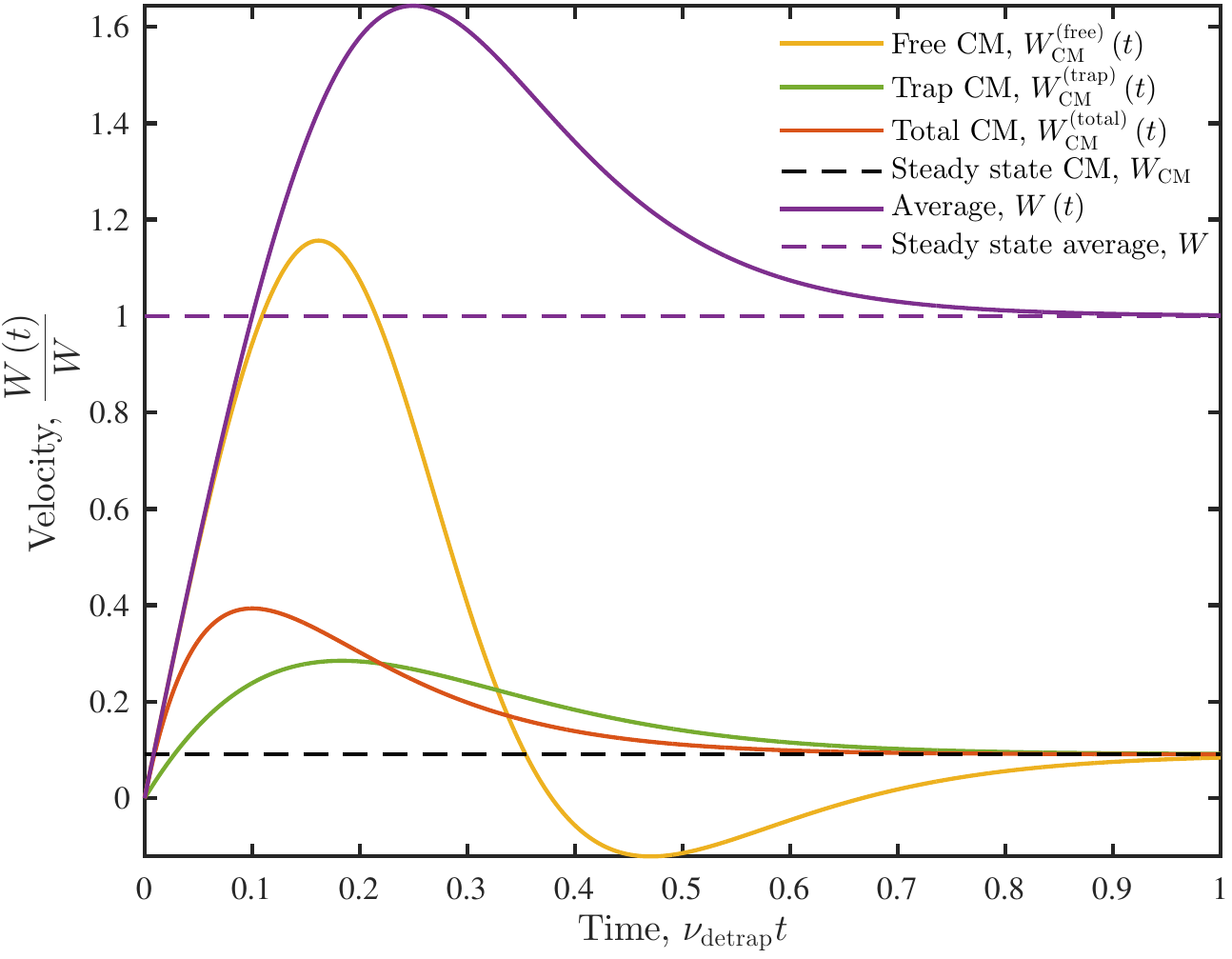}
\par\end{centering}

\centering{}\caption{\label{fig:negativeCoefficients}Plot of the centre of mass (CM) velocities
for free, trapped and total particles, $\mathbf{W}_{\mathrm{CM}}^{\left(\mathrm{free}\right)}\left(t\right)$,
$\mathbf{W}_{\mathrm{CM}}^{\left(\mathrm{trap}\right)}\left(t\right)$
and $\mathbf{W}_{\mathrm{CM}}^{\left(\mathrm{total}\right)}\left(t\right)$,
as well as the actual average velocity for the free particles, $\mathbf{W}\left(t\right)$,
for the exponential waiting time distribution $\phi\left(t\right)=\nu_{\mathrm{detrap}}\mathrm{e}^{-\nu_{\mathrm{detrap}}t}$.
The free particle CM velocity $\mathbf{W}_{\mathrm{CM}}^{\left(\mathrm{free}\right)}\left(t\right)$
and average velocity $\mathbf{W}\left(t\right)$ coincide linearly
at early times according to Eq. (\ref{eq:linearW}). The free particle
CM velocity $\mathbf{W}_{\mathrm{CM}}\left(t\right)$ is seen to transiently
become negative due to particles trapped early near the origin leaving
their traps. In this case, there is no recombination present, $\nu_{\mathrm{loss}}^{\left(\mathrm{free}\right)}=\nu_{\mathrm{loss}}^{\left(\mathrm{trap}\right)}=0$,
the collision frequency is set to $\nu_{\mathrm{coll}}/\nu_{\mathrm{detrap}}=1$
and the trapping rate is made sufficiently large so as the transient
negative velocity manifests, $\nu_{\mathrm{trap}}/\nu_{\mathrm{detrap}}=10$.
An additional consequence of this relatively large trapping rate is
that almost all free particles become trapped early on, allowing the
velocities to almost reach their steady state values after only a
single trapping time $\nu_{\mathrm{detrap}}t=1$.}
\end{figure}

The movement of the free particles can also be described by looking
directly at velocity moments of the generalised Boltzmann equation
(\ref{eq:boltzmannEquation})
\begin{eqnarray}
\mathcal{L}\left\{ \frac{N\left(t\right)}{N\left(0\right)}\left\langle \mathbf{v}\right\rangle \left(t\right)\right\}  & = & \frac{\mathbf{a}}{\tilde{p}\left(\tilde{p}+\tilde{\nu}\right)},\label{eq:avgVelocity}\\
\mathcal{L}\left\{ \frac{N\left(t\right)}{N\left(0\right)}\left\langle \mathbf{r}\mathbf{v}\right\rangle \left(t\right)\right\}  & = & \frac{\mathbf{I}}{\tilde{p}\left(\tilde{p}+\tilde{\nu}\right)^{2}}\left[\frac{\tilde{p}}{\alpha_{0}^{2}}+\frac{\nu_{\mathrm{coll}}}{\alpha_{\mathrm{coll}}^{2}}+\frac{\nu_{\mathrm{trap}}\Phi\left(p\right)}{\alpha_{\mathrm{detrap}}^{2}}\right]+\frac{\mathbf{a}\mathbf{a}}{\tilde{p}\left(\tilde{p}+\tilde{\nu}\right)^{2}}\left(\frac{1}{\tilde{p}}+\frac{2}{\tilde{p}+\tilde{\nu}}\right),
\end{eqnarray}
from which we define the average velocity
\begin{equation}
\mathbf{W}\left(t\right)\equiv\left\langle \mathbf{v}\right\rangle \left(t\right),\label{eq:Wavg}
\end{equation}
and average diffusivity
\begin{equation}
\mathbf{D}\left(t\right)\equiv\left\langle \mathbf{r}\mathbf{v}\right\rangle \left(t\right)-\left\langle \mathbf{r}\right\rangle \left(t\right)\left\langle \mathbf{v}\right\rangle \left(t\right).\label{eq:Davg}
\end{equation}
Figure \ref{fig:negativeCoefficients} plots the CM velocity $\mathbf{W}_{\mathrm{CM}}\left(t\right)$
for the free, trapped and total particles alongside the average velocity
$\mathbf{W}\left(t\right)$ for the free particles. We can see that
all measures of velocity begin at zero due to the Maxwellian initial
condition (\ref{eq:initialCondition}) being spherically symmetric
in velocity space. All velocities then increase due to the applied
field, with the free particle CM velocity $\mathbf{W}_{\mathrm{CM}}^{\left(\mathrm{free}\right)}\left(t\right)$
and average velocity $\mathbf{W}\left(t\right)$ coinciding linearly
at early times
\begin{equation}
\mathbf{W}\left(t\right)\approx\mathbf{W}_{\mathrm{CM}}^{\left(\mathrm{free}\right)}\left(t\right)\approx\mathbf{a}t.\label{eq:linearW}
\end{equation}
A similar small time expansion can be written for the free particle
diffusivites
\begin{equation}
\mathbf{D}\left(t\right)\approx\mathbf{D}_{\mathrm{CM}}^{\left(\mathrm{free}\right)}\left(t\right)\approx\frac{\mathbf{I}}{\alpha_{0}^{2}}t+\frac{\mathbf{a}\mathbf{a}}{12}\left[\nu_{\mathrm{coll}}+6\left(\nu_{\mathrm{loss}}^{\left(\mathrm{free}\right)}+\nu_{\mathrm{trap}}\right)\right]t^{4}.
\end{equation}
This coincidence between the free particle CM and average velocities
only lasts temporarily before the CM velocity decreases, becoming
negative prior to reaching its positive steady state value. This movement
of the free particle CM against the field is due to the processes
of trapping and detrapping. Specifically, as all particles are initially
untrapped, an unusually large \textquotedbl{}pulse\textquotedbl{}
of particles are trapped near the origin, which is later released,
causing a bias of the distribution and shifting the CM towards the
origin. Similarly, as the diffusivity of particles trapped early is
initially small, the free particle CM diffusivity $\mathbf{D}_{\mathrm{CM}}^{\left(\mathrm{free}\right)}\left(t\right)$
can also become transiently negative, as the distribution appears
to \textquotedbl{}bunch up\textquotedbl{} near the origin as the initial
pulse is released. Finally, we can see that all CM velocities approach
the same steady state value, while the free particle average velocity
approaches a separate steady state. Specifically, the CM transport
coefficients, $\mathbf{W}_{\mathrm{CM}}\left(t\right)$ and $\mathbf{D}_{\mathrm{CM}}\left(t\right)$,
approach the values given by Eqs. (\ref{eq:WCMlimit}) and (\ref{eq:DCMlimit}),
while the average transport coefficients, $\mathbf{W}\left(t\right)$
and $\mathbf{D}\left(t\right)$, approach Eqs. (\ref{eq:driftVelocity})
and (\ref{eq:diffusionCoefficient}).

\section{\label{sec:Hydrodynamic}Hydrodynamic regime and the generalised
diffusion equation}

\subsection{Chapman-Enskog perturbative solution}

The Chapman-Enskog perturbative solution technique \citep{Chapman1970}
assumes that certain terms in the Boltzmann equation are small relative
to others, allowing the solution to be written in the form of a Maclaurin
series expansion. Traditionally, the Chapman-Enskog expansion assumes
that both the explicit and implicit time derivatives in the Boltzmann
equation are small. An implication of this is that the perturbative
solution is valid only when the applied field is small. We will relax
this condition and instead use a generalisation of the Chapman-Enskog
expansion that only considers small explicit time and space derivatives,
known as a hydrodynamic expansion. We expect the resulting solution
to be most accurate in the long distance steady state. Note, however,
in Subsection \ref{sub:particleNumber} we determined that a steady
state is not always attainable for our generalised Boltzmann equation
(\ref{eq:boltzmannEquation}). Specifically, if there is \textit{any}
recombination present, all free and trapped particles are eventually
lost. To enforce that a steady state is always reached, we introduce
the scaled phase-space distribution function of constant particle
number $N\left(0\right)$
\begin{eqnarray}
F\left(t,\mathbf{r},\mathbf{v}\right) & \equiv & \frac{N\left(0\right)}{N\left(t\right)}f\left(t,\mathbf{r},\mathbf{v}\right).\label{eq:distributionRescale}
\end{eqnarray}
Substitution into the generalised Boltzmann equation (\ref{eq:boltzmannEquation})
provides a corresponding generalised Boltzmann equation for this scaled
distribution 
\begin{multline}
\left(\frac{\partial}{\partial t}+\mathbf{v}\cdot\frac{\partial}{\partial\mathbf{r}}+\mathbf{a}\cdot\frac{\partial}{\partial\mathbf{v}}\right)F\left(t,\mathbf{r},\mathbf{v}\right)=-\nu_{\mathrm{coll}}\left[F\left(t,\mathbf{r},\mathbf{v}\right)-n_{F}\left(t,\mathbf{r}\right)w\left(\alpha_{\mathrm{coll}},v\right)\right]\\
-\nu_{\mathrm{trap}}\left[R\left(t\right)F\left(t,\mathbf{r},\mathbf{v}\right)-R\left(t,\mathbf{r}\right)n_{F}\left(t,\mathbf{r}\right)w\left(\alpha_{\mathrm{detrap}},v\right)\right],
\end{multline}
where $n_{F}\left(t,\mathbf{r}\right)\equiv\int\mathrm{d}\mathbf{v}F\left(t,\mathbf{r},\mathbf{v}\right)$
and we have introduced the ratio of detrapping and trapping rates
\begin{equation}
R\left(t,\mathbf{r}\right)\equiv\frac{\Phi\left(t\right)\ast n\left(t,\mathbf{r}\right)}{n\left(t,\mathbf{r}\right)},\label{eq:R(t,r)}
\end{equation}
and its spatially homogeneous form
\begin{equation}
R\left(t\right)\equiv\frac{\Phi\left(t\right)\ast N\left(t\right)}{N\left(t\right)}.
\end{equation}
On the terms we wish to denote as small, we will temporarily introduce
a multiplicative parameter $\delta$
\begin{multline}
\delta\left(\frac{\partial}{\partial t}+\mathbf{v}\cdot\frac{\partial}{\partial\mathbf{r}}\right)F_{\delta}\left(t,\mathbf{r},\mathbf{v}\right)+\mathbf{a}\cdot\frac{\partial}{\partial\mathbf{v}}F_{\delta}\left(t,\mathbf{r},\mathbf{v}\right)=-\nu_{\mathrm{coll}}\left[F_{\delta}\left(t,\mathbf{r},\mathbf{v}\right)-n_{F}\left(t,\mathbf{r}\right)w\left(\alpha_{\mathrm{coll}},v\right)\right]\\
-\nu_{\mathrm{trap}}\left[R\left(t\right)F_{\delta}\left(t,\mathbf{r},\mathbf{v}\right)-R\left(t,\mathbf{r}\right)n_{F}\left(t,\mathbf{r}\right)w\left(\alpha_{\mathrm{detrap}},v\right)\right],\label{eq:deltaBoltzmann}
\end{multline}
through which we can expand the solution in a power series
\begin{equation}
F_{\delta}\left(t,\mathbf{r},\mathbf{v}\right)=\sum_{n\geq0}F^{\left(n\right)}\left(t,\mathbf{r},\mathbf{v}\right)\delta^{n}.\label{eq:deltaExpansion}
\end{equation}
This allows the actual solution to be recovered by setting $\delta=1$
in the above series expansion 
\begin{equation}
F\left(t,\mathbf{r},\mathbf{v}\right)=\sum_{n\geq0}F^{\left(n\right)}\left(t,\mathbf{r},\mathbf{v}\right).\label{eq:seriesSolution}
\end{equation}
The terms in this series solution can be found recursively by substituting
the $\delta$ expansion (\ref{eq:deltaExpansion}) for $F_{\delta}\left(t,\mathbf{r},\mathbf{v}\right)$
into the generalised Boltzmann equation (\ref{eq:deltaBoltzmann})
and equating powers of $\delta$
\begin{equation}
\left[\nu_{\mathrm{coll}}+\nu_{\mathrm{trap}}R\left(t\right)+\mathbf{a}\cdot\frac{\partial}{\partial\mathbf{v}}\right]F^{\left(n\right)}\left(t,\mathbf{r},\mathbf{v}\right)=-\left(\frac{\partial}{\partial t}+\mathbf{v}\cdot\frac{\partial}{\partial\mathbf{r}}\right)F^{\left(n-1\right)}\left(t,\mathbf{r},\mathbf{v}\right).\label{eq:chapmanRecurrence}
\end{equation}
This recurrence relationship is valid for $n\geq1$, with the initial
term given separately as
\begin{equation}
\left[1+\left\langle \mathbf{v}\right\rangle ^{\left(0\right)}\left(t\right)\cdot\frac{\partial}{\partial\mathbf{v}}\right]F^{\left(0\right)}\left(t,\mathbf{r},\mathbf{v}\right)=\frac{\nu_{\mathrm{coll}}w\left(\alpha_{\mathrm{coll}},v\right)+\nu_{\mathrm{trap}}R\left(t,\mathbf{r}\right)w\left(\alpha_{\mathrm{detrap}},v\right)}{\nu_{\mathrm{coll}}+\nu_{\mathrm{trap}}R\left(t,\mathbf{r}\right)}n_{F}\left(t,\mathbf{r}\right),
\end{equation}
in terms of its average velocity
\begin{equation}
\left\langle \mathbf{v}\right\rangle ^{\left(0\right)}\left(t\right)\equiv\frac{\mathbf{a}}{\nu_{\mathrm{coll}}+\nu_{\mathrm{trap}}R\left(t\right)}.
\end{equation}
Note here we have enforced the normalisation condition
\begin{equation}
\int\mathrm{d}\mathbf{v}F^{\left(0\right)}\left(t,\mathbf{r},\mathbf{v}\right)\equiv n_{F}\left(t,\mathbf{r}\right).
\end{equation}
In Fourier-transformed velocity space we can write this initial term
explicitly
\begin{equation}
F^{\left(0\right)}\left(t,\mathbf{r},\mathbf{s}\right)=\tilde{w}\left(t,\mathbf{r},\mathbf{s}\right)n_{F}\left(t,\mathbf{r}\right),
\end{equation}
where
\begin{equation}
\tilde{w}\left(t,\mathbf{r},\mathbf{s}\right)\equiv\frac{1}{1+\left\langle \mathbf{v}\right\rangle ^{\left(0\right)}\left(t\right)\cdot\imath\mathbf{s}}\frac{\nu_{\mathrm{coll}}w\left(\alpha_{\mathrm{coll}},s\right)+\nu_{\mathrm{trap}}R\left(t,\mathbf{r}\right)w\left(\alpha_{\mathrm{detrap}},s\right)}{\nu_{\mathrm{coll}}+\nu_{\mathrm{trap}}R\left(t,\mathbf{r}\right)}.\label{eq:wtrs}
\end{equation}
We can confirm that this approximate hydrodynamic solution is most
accurate in the steady state by noting that its average velocity coincides
with the actual average velocity (\ref{eq:Wavg}) at late times, $\lim_{t\rightarrow\infty}\left\langle \mathbf{v}\right\rangle ^{\left(0\right)}\left(t\right)=\lim_{t\rightarrow\infty}\left\langle \mathbf{v}\right\rangle \left(t\right)$.
We will denote this shared steady state velocity as 
\begin{equation}
\mathbf{W}\equiv\frac{\mathbf{a}}{\nu_{\mathrm{eff}}},\label{eq:driftVelocity}
\end{equation}
where the separate collision and trapping processes contribute to
the effective frequency
\begin{equation}
\nu_{\mathrm{eff}}\equiv\nu_{\mathrm{coll}}+R\nu_{\mathrm{trap}},\label{eq:effectiveFrequency}
\end{equation}
defined in terms of the spatially averaged limiting ratio of detrapping
and trapping rates
\begin{equation}
R\equiv\lim_{t\rightarrow\infty}R\left(t\right).\label{eq:Rdefinition}
\end{equation}
This limit can be evaluated implicitly as satisfying
\begin{equation}
R\equiv\int_{0}^{\infty}\mathrm{d}t\Phi\left(t\right)\mathrm{e}^{\left[\nu_{\mathrm{loss}}^{\left(\mathrm{free}\right)}+\nu_{\mathrm{trap}}\left(1-R\right)\right]t}.\label{eq:Rrelation}
\end{equation}
Specific expressions for $R$ for various choices of the waiting time
distribution $\phi\left(t\right)$ are listed in Appendix \ref{sec:R}.
In terms of this velocity, $\mathbf{W}$, we can write the steady
state limit of Eq. (\ref{eq:wtrs}) as
\begin{equation}
\tilde{w}\left(\mathbf{r},\mathbf{s}\right)=\frac{1}{1+\mathbf{W}\cdot\imath\mathbf{s}}\frac{\nu_{\mathrm{coll}}w\left(\alpha_{\mathrm{coll}},s\right)+\nu_{\mathrm{trap}}R\left(\mathbf{r}\right)w\left(\alpha_{\mathrm{detrap}},s\right)}{\nu_{\mathrm{coll}}+\nu_{\mathrm{trap}}R\left(\mathbf{r}\right)},
\end{equation}
where the limiting ratio of detrapping and trapping rates is
\begin{equation}
R\left(\mathbf{r}\right)\equiv\lim_{t\rightarrow\infty}R\left(t,\mathbf{r}\right).
\end{equation}
In direct analogy with the implicit definition (\ref{eq:Rrelation})
of $R$ we have the following implicit definition of $R\left(\mathbf{r}\right)$
\begin{equation}
R\left(\mathbf{r}\right)\equiv\int_{0}^{\infty}\mathrm{d}t\Phi\left(t\right)\exp\left\{ \left[\nu_{\mathrm{loss}}^{\left(\mathrm{free}\right)}+\nu_{\mathrm{trap}}\left(1-R\left(\mathbf{r}\right)\right)+\frac{1}{n}\frac{\partial}{\partial\mathbf{r}}\cdot n\left\langle \mathbf{v}\right\rangle \right]t\right\} .\label{eq:Rrrelation}
\end{equation}
Finally, we can explore the spatial dependence of $\tilde{w}\left(\mathbf{r},\mathbf{s}\right)$
by considering a perturbation from its spatially averaged state
\begin{equation}
\tilde{w}\left(\mathbf{s}\right)=\frac{\nu_{\mathrm{coll}}w\left(\alpha_{\mathrm{coll}},s\right)+\nu_{\mathrm{trap}}Rw\left(\alpha_{\mathrm{detrap}},s\right)}{\nu_{\mathrm{coll}}+\nu_{\mathrm{trap}}R+\mathbf{a}\cdot\imath\mathbf{s}}.
\end{equation}
To spatially perturb $\tilde{w}\left(\mathbf{s}\right)$, we must
first spatially perturb $R$ using the definition (\ref{eq:Rrrelation})
of $R\left(\mathbf{r}\right)$. Introducing the first order spatial
perturbation $\delta R$ and using the asymptotic velocity in the
hydrodynamic regime, $\left\langle \mathbf{v}\right\rangle \sim\mathbf{W}$,
provides the expression
\begin{equation}
R+\delta R=R\left\langle \exp\left[\left(\frac{1}{n}\frac{\partial n}{\partial\mathbf{r}}\cdot\mathbf{W}-\nu_{\mathrm{trap}}\delta R\right)t\right]\right\rangle ,
\end{equation}
in terms of the time average defined by
\begin{equation}
\left\langle \eta\left(t\right)\right\rangle \equiv\frac{1}{R}\int_{0}^{\infty}\mathrm{d}t\Phi\left(t\right)\mathrm{e}^{\left[\nu_{\mathrm{loss}}^{\left(\mathrm{free}\right)}+\nu_{\mathrm{trap}}\left(1-R\right)\right]t}\eta\left(t\right).\label{eq:timeAverages}
\end{equation}
Performing a power series expansion and truncating beyond first order
gives the solution $R\left(\mathbf{r}\right)=R+\delta R$ as a density
gradient expansion up to first order
\begin{equation}
R\left(\mathbf{r}\right)=R+\mathbf{R}^{\left(1\right)}\cdot\frac{1}{n}\frac{\partial n}{\partial\mathbf{r}},\label{eq:spatialR}
\end{equation}
in terms of the vector coefficient
\begin{equation}
\mathbf{R}^{\left(1\right)}\equiv\frac{R\left\langle t\right\rangle }{1+\nu_{\mathrm{trap}}R\left\langle t\right\rangle }\mathbf{W}.
\end{equation}
Now the spatially averaged steady state velocity distribution $\tilde{w}\left(\mathbf{s}\right)$
can be spatially perturbed using the density gradient expansion (\ref{eq:spatialR}),
resulting in, to first spatial order
\begin{eqnarray}
F^{\left(0\right)}\left(t,\mathbf{r},\mathbf{s}\right) & = & \tilde{w}\left(\mathbf{s}\right)n_{F}\left(t,\mathbf{r}\right)\nonumber \\
 &  & +\frac{w\left(\alpha_{\mathrm{detrap}},s\right)-\tilde{w}\left(\mathbf{s}\right)}{\nu_{\mathrm{coll}}+\nu_{\mathrm{trap}}R+\mathbf{a}\cdot\imath\mathbf{s}}\nu_{\mathrm{trap}}\mathbf{R}^{\left(1\right)}\cdot\frac{\partial}{\partial\mathbf{r}}n_{F}\left(t,\mathbf{r}\right).
\end{eqnarray}
Using the recurrence relationship (\ref{eq:chapmanRecurrence}) and
the continuity equation (\ref{eq:particleBalance}) to evaluate the
explicit time derivative provides the next term, also to first spatial
order
\begin{eqnarray}
F^{\left(1\right)}\left(t,\mathbf{r},\mathbf{s}\right) & = & \frac{\mathbf{W}\tilde{w}\left(\mathbf{s}\right)-\imath\frac{\partial}{\partial\mathbf{s}}\tilde{w}\left(\mathbf{s}\right)}{\nu_{\mathrm{coll}}+\nu_{\mathrm{trap}}R+\mathbf{a}\cdot\imath\mathbf{s}}\cdot\frac{\partial}{\partial\mathbf{r}}n_{F}\left(t,\mathbf{r}\right).
\end{eqnarray}
Similarly, $F^{\left(2\right)}\left(t,\mathbf{r},\mathbf{v}\right)$
can be found and shown to be of minimum second order in spatial gradients.
In general, $F^{\left(n\right)}\left(t,\mathbf{r},\mathbf{v}\right)$
is described by a full density gradient expansion of minimum spatial
order $n$. Including all zeroth and first order contributions, the
generalised Boltzmann equation solution is
\begin{eqnarray}
f\left(t,\mathbf{r},\mathbf{s}\right) & = & \tilde{w}\left(\mathbf{s}\right)n\left(t,\mathbf{r}\right)\nonumber \\
 &  & +\frac{\left[\mathbf{W}-\nu_{\mathrm{trap}}\mathbf{R}^{\left(1\right)}-\imath\frac{\partial}{\partial\mathbf{s}}\right]\tilde{w}\left(\mathbf{s}\right)+\nu_{\mathrm{trap}}\mathbf{R}^{\left(1\right)}w\left(\alpha_{\mathrm{detrap}},s\right)}{\nu_{\mathrm{coll}}+\nu_{\mathrm{trap}}R+\mathbf{a}\cdot\imath\mathbf{s}}\cdot\frac{\partial}{\partial\mathbf{r}}n\left(t,\mathbf{r}\right).\label{eq:fDensity}
\end{eqnarray}
Velocity integration provides Fick's law for the free particle flux
\begin{equation}
n\left\langle \mathbf{v}\right\rangle =\mathbf{W}n-\mathbf{D}\cdot\frac{\partial n}{\partial\mathbf{r}},\label{eq:fluxDensity}
\end{equation}
which implies that $\mathbf{W}$ is the flux drift velocity and defines
the flux diffusion coefficient as
\begin{equation}
\mathbf{D}\equiv\frac{1}{\nu_{\mathrm{eff}}}\left[\frac{\mathbf{I}}{\alpha_{\mathrm{eff}}^{2}}+\left(\mathbf{W}+\nu_{\mathrm{trap}}\mathbf{R}^{\left(1\right)}\right)\mathbf{W}\right],\label{eq:diffusionCoefficient}
\end{equation}
written in terms of the effective frequency (\ref{eq:effectiveFrequency})
and the effective temperature
\begin{eqnarray}
T_{\mathrm{eff}} & \equiv & \frac{\nu_{\mathrm{coll}}}{\nu_{\mathrm{coll}}+R\nu_{\mathrm{trap}}}T_{\mathrm{coll}}+\frac{R\nu_{\mathrm{trap}}}{\nu_{\mathrm{coll}}+R\nu_{\mathrm{trap}}}T_{\mathrm{detrap}}.
\end{eqnarray}
Similar to the flux drift velocity $\mathbf{W}$, the flux diffusion
coefficient $\mathbf{D}$ could have also been derived as the long
time limit of the average diffusivity (\ref{eq:Davg})
\begin{equation}
\mathbf{D}\equiv\lim_{t\rightarrow\infty}\left[\left\langle \mathbf{r}\mathbf{v}\right\rangle \left(t\right)-\left\langle \mathbf{r}\right\rangle \left(t\right)\left\langle \mathbf{v}\right\rangle \left(t\right)\right].
\end{equation}
The flux diffusion coefficient derived here differs slightly from
what was derived in \citep{Philippa2014} for a similar phase-space
kinetic model utilising the same operator for trapping and detrapping.
It is likely they did not consider the spatial dependence in Eq. (\ref{eq:R(t,r)})
for the ratio of detrapping and trapping rates $R\left(t,\mathbf{r}\right)$,
as their diffusion coefficient lacked the additional anisotropic component
$\frac{\nu_{\mathrm{trap}}}{\nu_{\mathrm{eff}}}\mathbf{R}^{\left(1\right)}\mathbf{W}$.
Subsequently, their diffusion coefficient is only valid in the isotropic
case without an applied field, where $\mathbf{W}=0$, or in the limit
of instantaneous detrapping, where $\mathbf{R}^{\left(1\right)}=\mathbf{0}$.

\subsection{Analytical correspondence of transport coefficients}

\subsubsection{Diffusion equations in the hydrodynamic regime}

In the previous subsection we considered a perturbative solution of
the generalised Boltzmann equation (\ref{eq:boltzmannEquation}),
written in the hydrodynamic regime as the density gradient expansion
(\ref{eq:fDensity}). This solution directly provided the flux transport
coefficients of velocity (\ref{eq:driftVelocity}) and diffusion (\ref{eq:diffusionCoefficient}).
In this subsection, we look to reconcile these results analytically
using Eq. (\ref{eq:continuitySolution}) for the number density. We
can describe the asymptotics of the number density by looking at its
poles in Laplace space, given by solving the dispersion relation \citep{Robson1975}
\begin{equation}
1-\nu_{\mathrm{coll}}\zeta_{\mathrm{coll}}\left(\mathbf{k}\right)Z\left[-\left(\tilde{p}+\tilde{\nu}\right)\zeta_{\mathrm{coll}}\left(\mathbf{k}\right)\right]-\nu_{\mathrm{trap}}\Phi\left(p\right)\zeta_{\mathrm{detrap}}\left(\mathbf{k}\right)Z\left[-\left(\tilde{p}+\tilde{\nu}\right)\zeta_{\mathrm{detrap}}\left(\mathbf{k}\right)\right]=0.
\end{equation}
Using the asymptotic series representation of the plasma dispersion
function \citep{fried1961plasma}
\begin{equation}
Z\left(\xi\right)=-\frac{1}{\xi}\sum_{n\geq0}\frac{\left(2n-1\right)!!}{2^{n}}\xi^{-2n}=-\frac{1}{\xi}\left(1+\frac{1}{2}\xi^{-2}+\frac{3}{4}\xi^{-4}+\cdots\right),
\end{equation}
we perform a small $\mathbf{k}$ expansion and find the root of the
dispersion relation to second spatial order
\begin{equation}
p=-\nu_{\mathrm{trap}}\left(1-R\right)-\nu_{\mathrm{loss}}^{\left(\mathrm{free}\right)}-\mathbf{W}_{\mathrm{CM}}\cdot\imath\mathbf{k}+\mathbf{D}_{\mathrm{CM}}\colon\imath\mathbf{k}\imath\mathbf{k},
\end{equation}
which corresponds to the diffusion equation
\begin{equation}
\left[\frac{\partial}{\partial t}+\nu_{\mathrm{trap}}\left(1-R\right)+\nu_{\mathrm{loss}}^{\left(\mathrm{free}\right)}+\mathbf{W}_{\mathrm{CM}}\cdot\frac{\partial}{\partial\mathbf{r}}-\mathbf{D}_{\mathrm{CM}}\colon\frac{\partial^{2}}{\partial\mathbf{r}\partial\mathbf{r}}\right]n\left(t,\mathbf{r}\right)=0.\label{eq:normalDiffusionCM}
\end{equation}
Here the steady state centre of mass (CM) transport coefficients are
defined
\begin{eqnarray}
\mathbf{W}_{\mathrm{CM}} & \equiv & \frac{\mathbf{R}^{\left(1\right)}}{R\left\langle t\right\rangle },\label{eq:WCMlimit}\\
\mathbf{D}_{\mathrm{CM}} & \equiv & \frac{\left\langle t^{2}\right\rangle }{2\left\langle t\right\rangle }\mathbf{W}_{\mathrm{CM}}\mathbf{W}_{\mathrm{CM}}-\frac{\mathbf{R}^{\left(2\right)}}{R\left\langle t\right\rangle },\label{eq:DCMlimit}
\end{eqnarray}
using the density gradient expansion of $R\left(\mathbf{r}\right)$
\begin{equation}
R\left(\mathbf{r}\right)=R+\mathbf{R}^{\left(1\right)}\cdot\frac{1}{n}\frac{\partial n}{\partial\mathbf{r}}+\mathbf{R}^{\left(2\right)}\colon\frac{1}{n}\frac{\partial^{2}n}{\partial\mathbf{r}\partial\mathbf{r}},
\end{equation}
written now to second order using the flux diffusion coefficient $\mathbf{D}$
\begin{equation}
\mathbf{R}^{\left(2\right)}\equiv\frac{R\left\langle t^{2}\right\rangle }{2\left(1+\nu_{\mathrm{trap}}R\left\langle t\right\rangle \right)^{3}}\mathbf{W}\mathbf{W}-\frac{R\left\langle t\right\rangle }{1+\nu_{\mathrm{trap}}R\left\langle t\right\rangle }\mathbf{D},
\end{equation}
where time averages are defined by Eq. (\ref{eq:timeAverages}). Substitution
of the root of the dispersion relation into the time operator of the
continuity equation yields, to second spatial order
\begin{equation}
p+\nu_{\mathrm{trap}}\left[1-\Phi\left(p\right)\right]+\nu_{\mathrm{loss}}^{\left(\mathrm{free}\right)}=-\mathbf{W}\cdot\imath\mathbf{k}+\mathbf{D}\colon\imath\mathbf{k}\imath\mathbf{k},
\end{equation}
which corresponds to the generalised diffusion equation
\begin{equation}
\left[\frac{\partial}{\partial t}+\nu_{\mathrm{trap}}\left(1-\Phi\left(t\right)\ast\right)+\nu_{\mathrm{loss}}^{\left(\mathrm{free}\right)}+\mathbf{W}\cdot\frac{\partial}{\partial\mathbf{r}}-\mathbf{D}\colon\frac{\partial^{2}}{\partial\mathbf{r}\partial\mathbf{r}}\right]n\left(t,\mathbf{r}\right)=0,\label{eq:generalisedDiffusion}
\end{equation}
in terms of the flux transport coefficients $\mathbf{W}$ and $\mathbf{D}$.
This could have alternatively been derived by approximating the flux
in the continuity equation directly using its density gradient expansion
(Fick's law) given by Eq. (\ref{eq:fluxDensity}).

\subsubsection{Approaching the steady state}

So far we have considered the continuity equation in both the steady
and near spatially homogeneous state. Using the analytical solution,
it is possible to relax this steady state assumption. We can write
the flux exactly by rearranging the continuity equation (\ref{eq:particleBalance})
in Fourier-Laplace space
\begin{equation}
\imath\mathbf{k}\cdot\mathcal{L}\left\{ n\left(t,\mathbf{k}\right)\left\langle \mathbf{v}\right\rangle \left(t\right)\right\} =\left[\frac{N\left(0\right)}{n\left(p,\mathbf{k}\right)}-\tilde{p}\right]n\left(p,\mathbf{k}\right).
\end{equation}
Performing a small $\mathbf{k}$ expansion of the above coefficient
of $n\left(p,\mathbf{k}\right)$ gives an approximate continuity equation
valid for large distances or near spatially homogeneous states
\begin{equation}
\left[p+\nu_{\mathrm{trap}}\left(1-\Phi\left(p\right)\right)+\nu_{\mathrm{loss}}^{\left(\mathrm{free}\right)}+\boldsymbol{\mathfrak{W}}\left(p\right)\cdot\imath\mathbf{k}-\boldsymbol{\mathfrak{D}}\left(p\right)\colon\imath\mathbf{k}\imath\mathbf{k}\right]n\left(p,\mathbf{k}\right)=N\left(0\right),
\end{equation}
where the following $p$-dependent velocity and diffusivity are defined
in Laplace space
\begin{eqnarray}
\boldsymbol{\mathfrak{W}}\left(p\right) & \equiv & \frac{\mathbf{a}}{\tilde{p}+\tilde{\nu}},\\
\boldsymbol{\mathfrak{D}}\left(p\right) & \equiv & \frac{\mathbf{I}}{\left(\tilde{p}+\tilde{\nu}\right)^{2}}\left[\frac{\tilde{p}}{\alpha_{0}^{2}}+\frac{\nu_{\mathrm{coll}}}{\alpha_{\mathrm{coll}}^{2}}+\frac{\nu_{\mathrm{trap}}\Phi\left(p\right)}{\alpha_{\mathrm{detrap}}^{2}}\right]+\frac{2\mathbf{a}\mathbf{a}}{\left(\tilde{p}+\tilde{\nu}\right)^{3}},
\end{eqnarray}
although in the time domain $\boldsymbol{\mathfrak{W}}\left(t\right)$
and $\boldsymbol{\mathfrak{D}}\left(t\right)$ have units of length
and area respectively. Performing the inverse Fourier-Laplace transform
yields
\begin{equation}
\left[\frac{\partial}{\partial t}+\nu_{\mathrm{trap}}\left(1-\Phi\left(t\right)\ast\right)+\nu_{\mathrm{loss}}^{\left(\mathrm{free}\right)}+\boldsymbol{\mathfrak{W}}\left(t\right)\ast\cdot\frac{\partial}{\partial\mathbf{r}}-\boldsymbol{\mathfrak{D}}\left(t\right)\ast\colon\frac{\partial^{2}}{\partial\mathbf{r}\partial\mathbf{r}}\right]n\left(t,\mathbf{r}\right)=0,
\end{equation}
which is of a similar form to the generalised diffusion equation,
but now the ``transport coefficients'' are time convolved with the
number density. It should be noted that, as the flux has been written
to second spatial order, the first and second order spatial moments
of this approximate continuity equation are exact for all times.

\section{\label{sec:Fractional}Connection with fractional transport}

Dispersive transport is physically characterised by long-lived traps
\citep{scher1975anomalous}. For the right choice of parameters, the
generalised Boltzmann equation (\ref{eq:boltzmannEquation}) is capable
of modelling such trapped states. A necessary condition for dispersive
transport is a waiting time distribution with a divergent mean. One
choice is a waiting time distribution with a heavy tail of the power
law form
\begin{equation}
\phi\left(t\right)\sim t^{-\left(1+\alpha\right)},
\end{equation}
where $0<\alpha<1$. This takes the small $p$ form in Laplace space
\begin{equation}
\phi\left(p\right)\approx1-r_{\alpha}p^{\alpha}.
\end{equation}
Additionally, we must enforce that no trap-based recombination occurs,
$\nu_{\mathrm{loss}}^{\left(\mathrm{trap}\right)}=0$, as this has
the effect of prematurely shortening the trapping time so that the
mean trapping time no longer diverges. In this case, the effective
waiting time distribution (\ref{eq:effectivePhi}) is no longer weighted
by an exponential decay term, $\Phi\left(t\right)\rightarrow\phi\left(t\right)$,
and the continuity equation (\ref{eq:particleBalance}) becomes
\begin{equation}
\left[\frac{\partial}{\partial t}+\nu_{\mathrm{trap}}\left(1-\phi\left(t\right)\ast\right)+\nu_{\mathrm{loss}}^{\left(\mathrm{free}\right)}\right]n\left(t,\mathbf{r}\right)+\frac{\partial}{\partial\mathbf{r}}\cdot\left[n\left(t,\mathbf{r}\right)\left\langle \mathbf{v}\right\rangle \left(t,\mathbf{r}\right)\right]=0.
\end{equation}
We can separate the power law tail from the waiting time distribution
\begin{equation}
\phi\left(t\right)\ast n\left(t,\mathbf{r}\right)=\psi\left(t\right)\ast n\left(t,\mathbf{r}\right)-r_{\alpha}\left[_{0}^{\mathrm{C}}\mathcal{D}_{t}^{\alpha}n\left(t,\mathbf{r}\right)+\frac{t^{-\alpha}}{\Gamma\left(1-\alpha\right)}n\left(0,\mathbf{r}\right)\right],
\end{equation}
where the moments of $\psi\left(t\right)$ are well-defined and the
operator of Caputo fractional differentiation of order $\alpha$ is
defined
\begin{equation}
_{0}^{\mathrm{C}}\mathcal{D}_{t}^{\alpha}n\left(t,\mathbf{r}\right)\equiv\frac{1}{\Gamma\left(1-\alpha\right)}\int_{0}^{t}\mathrm{d}\tau\left(t-\tau\right)^{-\alpha}\frac{\partial}{\partial\tau}n\left(\tau,\mathbf{r}\right).
\end{equation}
The continuity equation can now be written exactly as
\begin{equation}
\left[\frac{\partial}{\partial t}+r_{\alpha}\nu_{\mathrm{trap}}{}_{0}^{\mathrm{C}}\mathcal{D}_{t}^{\alpha}+\nu_{\mathrm{trap}}\left(1-\psi\left(t\right)\ast\right)+\nu_{\mathrm{loss}}^{\left(\mathrm{free}\right)}\right]n\left(t,\mathbf{r}\right)+\frac{\partial}{\partial\mathbf{r}}\cdot\left[n\left(t,\mathbf{r}\right)\left\langle \mathbf{v}\right\rangle \left(t,\mathbf{r}\right)\right]=-\frac{r_{\alpha}\nu_{\mathrm{trap}}}{t^{\alpha}\Gamma\left(1-\alpha\right)}n\left(0,\mathbf{r}\right).
\end{equation}
Truncating the small $p$ expansion in Laplace space yields a form
of the continuity equation valid for long times
\begin{equation}
\left(_{0}^{\mathrm{C}}\mathcal{D}_{t}^{\alpha}+\frac{\nu_{\mathrm{loss}}^{\left(\mathrm{free}\right)}}{r_{\alpha}\nu_{\mathrm{trap}}}\right)n\left(t,\mathbf{r}\right)+\frac{\partial}{\partial\mathbf{r}}\cdot\left[n\left(t,\mathbf{r}\right)\frac{\left\langle \mathbf{v}\right\rangle \left(t,\mathbf{r}\right)}{r_{\alpha}\nu_{\mathrm{trap}}}\right]=\left[\delta\left(r_{\alpha}\nu_{\mathrm{trap}}t\right)-\frac{t^{-\alpha}}{\Gamma\left(1-\alpha\right)}\right]n\left(0,\mathbf{r}\right),
\end{equation}
written now solely in terms of the time operator of fractional differentiation.
Finally, performing a small $\mathbf{k}$ approximation in Fourier
space provides the Captuo time-fractional advection-diffusion equation
\begin{equation}
\left(_{0}^{\mathrm{C}}\mathcal{D}_{t}^{\alpha}+\frac{\nu_{\mathrm{loss}}^{\left(\mathrm{free}\right)}}{r_{\alpha}\nu_{\mathrm{trap}}}+\mathbf{W}_{\alpha}\cdot\frac{\partial}{\partial\mathbf{r}}-\mathbf{D}_{\alpha}\cdot\frac{\partial^{2}}{\partial\mathbf{r}\partial\mathbf{r}}\right)n\left(t,\mathbf{r}\right)=\left[\delta\left(r_{\alpha}\nu_{\mathrm{trap}}t\right)-\frac{t^{-\alpha}}{\Gamma\left(1-\alpha\right)}\right]n\left(0,\mathbf{r}\right),\label{eq:fractionalDiffusion}
\end{equation}
with fractional transport coefficients defined as
\begin{eqnarray}
\mathbf{W}_{\alpha} & \equiv & \frac{\mathbf{W}}{\nu_{\mathrm{trap}}r_{\alpha}},\\
\mathbf{D}_{\alpha} & \equiv & \frac{\mathbf{D}}{\nu_{\mathrm{trap}}r_{\alpha}},
\end{eqnarray}
in terms of the flux drift velocity (\ref{eq:driftVelocity}) and
diffusion coefficient (\ref{eq:diffusionCoefficient}), respectively.
Note that, as the waiting time distribution $\phi\left(t\right)$
has a divergent mean, the flux diffusion coefficient takes the particular
form
\begin{equation}
\mathbf{D}\equiv\frac{1}{\nu_{\mathrm{eff}}}\left(\frac{\mathbf{I}}{\alpha_{\mathrm{eff}}^{2}}+2\mathbf{W}\mathbf{W}\right).
\end{equation}
There exist generalisations of time-fractional diffusion equations,
like Eq. (\ref{eq:fractionalDiffusion}), where spatial derivatives
are also taken to be of non-integer order \citep{mainardi2007fundamental}.
Physically, these fractional space derivatives arise when particles
undergo long jumps in space \citep{metzler2004restaurant}. This is
analogous to the above situation where a time-fractional diffusion
equation arose from particles experiencing traps of long duration.
As our model currently only allows for variation in the trapping time,
we conclude that to similarly derive a space-fractional diffusion
equation would require adjustments to the kinetic theory.

Importantly, we should also note that a similar asymptotic approximation
of the generalised Boltzmann equation (\ref{eq:boltzmannEquation})
does not result in a fractional time operator that acts on the phase-space
distribution function. That is, it does not seem possible to derive
a similar ``fractional Boltzmann equation'' from our model. This
conclusion differs from \citep{Robson2005} who used a similar kinetic
model to successfully derive a fractional Boltzmann equation. However,
their model was inconsistent as it simultaneously modelled trapping
while also maintaining a constant number of free particles.

\section{\label{sec:Subordination}Subordination transformation}

As shown in the previous section, the generalised diffusion equation
is capable of describing dispersive transport in the same way the
Caputo fractional diffusion equation does. A general feature shared
by both of these diffusion equations is the history dependence of
their solutions. This is physically due to the existence of trapped
states and delayed detrapping. Mathematically, this manifests as a
global time operator, be it a fractional derivative or, in the case
of the generalised diffusion equation (\ref{eq:generalisedDiffusion}),
a convolution with the effective waiting time distribution $\Phi\left(t\right)$.

Due to their nature, global operators introduce additional complexity
when it comes to solving problems numerically. For example, in finite
difference schemes the computation time generally scales linearly
with the number of time steps chosen. The exception being when a global
time operator is present, causing the computation time to scale \textit{quadratically}
with the number of time steps. Although this increased computational
complexity is inherent to these systems, a number of techniques have
been suggested to improve upon it for fractional differential equations
\citep{podlubny1998fractional,Ford2001,fukunaga2011high,Stokes2015}.
One approach involves first solving a standard diffusion equation
and then performing a subordination integral transformation \citep{barkai2001fractional,Stokes2015}
to find the desired solution of a fractional diffusion equation. We
will generalise this approach to solve the generalised diffusion equation
for the free particle number density $n\left(t,\mathbf{r}\right)$.

Replacing the time operator in the generalised diffusion equation
with an explicit time derivative yields a standard diffusion equation
with the same linear spatial operator
\begin{equation}
\left(\frac{\partial}{\partial\tau}+\mathbf{W}\cdot\frac{\partial}{\partial\mathbf{r}}-\mathbf{D}\colon\frac{\partial^{2}}{\partial\mathbf{r}\partial\mathbf{r}}\right)u\left(\tau,\mathbf{r}\right)=0.\label{eq:normalDiffusion}
\end{equation}
For the same initial conditions, $u\left(0,\mathbf{r}\right)\equiv n\left(0,\mathbf{r}\right)$,
we can relate both solutions directly in Laplace space
\begin{equation}
n\left(p,\mathbf{r}\right)=u\left(p+\nu_{\mathrm{loss}}^{\left(\mathrm{free}\right)}+\nu_{\mathrm{trap}}\left[1-\Phi\left(p\right)\right],\mathbf{r}\right),
\end{equation}
which in the time domain corresponds to the subordination integral
transform
\begin{eqnarray}
n\left(t,\mathbf{r}\right) & \equiv & \mathcal{A}u\left(t,\mathbf{r}\right)\nonumber \\
 & \equiv & \int_{0}^{t}\mathrm{d}\tau A\left(\tau,t-\tau\right)u\left(\tau,\mathbf{r}\right),\label{eq:subordinationTransformation}
\end{eqnarray}
where the kernel is defined in terms of the inverse Laplace transform
$\mathcal{L}^{-1}$ 
\begin{equation}
A\left(\tau,t\right)\equiv\mathrm{e}^{-\left(\nu_{\mathrm{loss}}^{\left(\mathrm{free}\right)}+\nu_{\mathrm{trap}}\right)\tau}\mathrm{e}^{-\nu_{\mathrm{loss}}^{\left(\mathrm{trap}\right)}t}\mathcal{L}^{-1}\left\{ e^{\nu_{\mathrm{trap}}\phi\left(p\right)\tau}\right\} .\label{eq:integralKernel}
\end{equation}
Appendix \ref{sec:A} contains kernels corresponding to various choices
of the waiting time distribution $\phi\left(t\right)$.

\begin{figure}
\includegraphics{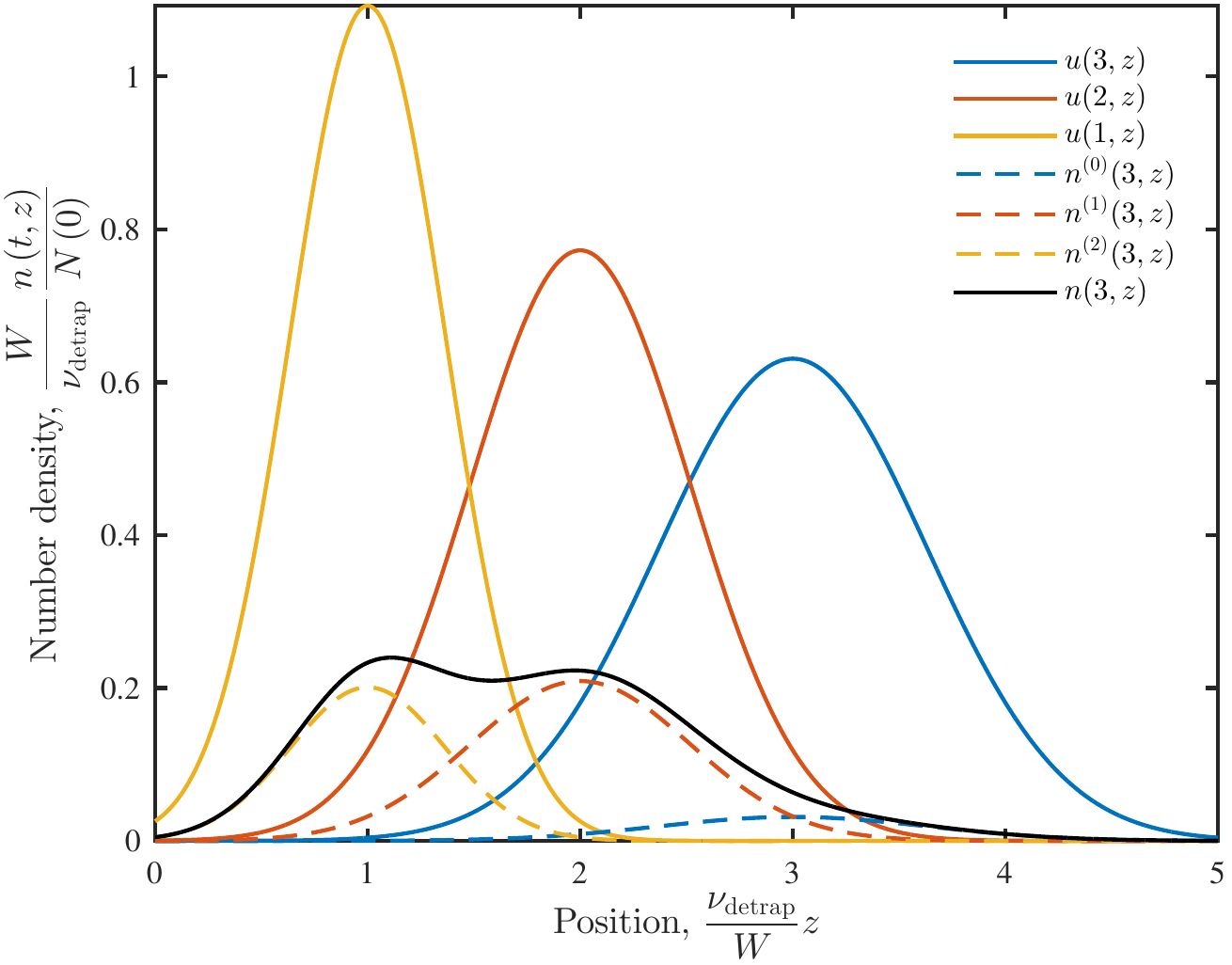}

\caption{\label{fig:subordinationExample}The solution $n\left(t,z\right)$
of the generalised diffusion equation (\ref{eq:generalisedDiffusion})
is written by sampling from the Gaussian solution $u\left(t,z\right)$
of the standard diffusion equation (\ref{eq:normalDiffusion}) at
multiple points in time. This is done using the subordination transformation
(\ref{eq:subordinationTransformation}). Here, traps are of fixed
duration, $\nu_{\mathrm{detrap}}t=1$, as described by the waiting
time distribution $\phi\left(t\right)/\nu_{\mathrm{detrap}}=\delta\left(\nu_{\mathrm{detrap}}t-1\right)$.
In this case, the subordination transformation becomes the summation
(\ref{eq:subordinationSummation}) whose individual terms $n^{\left(k\right)}\left(t,z\right)$
correspond to those free particles which have been trapped $k$ times
in the past. Here, there is no recombination, $\nu_{\mathrm{loss}}^{\left(\mathrm{free}\right)}=\nu_{\mathrm{loss}}^{\left(\mathrm{trap}\right)}=0$,
the trapping rate is set to $\nu_{\mathrm{trap}}/\nu_{\mathrm{detrap}}=1$
and the diffusion coefficient is made small so as to emphasise each
individual Gaussian's contribution to the solution, $D\nu_{\mathrm{detrap}}/W^{2}=1/15$.}
\end{figure}

As a simple example, consider the case of a shifted Dirac delta waiting
time distribution
\begin{equation}
\phi\left(t\right)=\delta\left(t-\nu_{\mathrm{detrap}}^{-1}\right),\label{eq:shiftedDirac}
\end{equation}
corresponding to traps of fixed duration $\nu_{\mathrm{detrap}}^{-1}$.
In this case, the subordination transformation (\ref{eq:subordinationTransformation})
simply becomes the summation
\begin{equation}
n\left(t,\mathbf{r}\right)=\sum_{k\geq0}n^{\left(k\right)}\left(t-k\nu_{\mathrm{detrap}}^{-1},\mathbf{r}\right),\label{eq:subordinationSummation}
\end{equation}
the terms of which can be physically interpreted as those free particles
which have been trapped $k$ times in the past
\begin{equation}
n^{\left(k\right)}\left(\tau,\mathbf{r}\right)\equiv H\left(\tau\right)\frac{\left(\mathrm{e}^{-\frac{\nu_{\mathrm{loss}}^{\left(\mathrm{trap}\right)}}{\nu_{\mathrm{detrap}}}}\nu_{\mathrm{trap}}\tau\right)^{k}}{k!}\mathrm{e}^{-\left(\nu_{\mathrm{loss}}^{\left(\mathrm{free}\right)}+\nu_{\mathrm{trap}}\right)\tau}u\left(\tau,\mathbf{r}\right),\label{eq:partialSolution}
\end{equation}
where $H\left(\tau\right)$ is the Heaviside step function. Figure
\ref{fig:subordinationExample} plots this solution on a one-dimensional
unbounded domain $z\in\left(-\infty,\infty\right)$ for the impulse
initial condition $n\left(0,z\right)\equiv N\left(0\right)\delta\left(z\right)$,
and shows its construction in terms of the corresponding Gaussian
solution of the standard diffusion equation (\ref{eq:normalDiffusion})
\begin{equation}
u\left(t,z\right)=\frac{N\left(0\right)}{2\sqrt{\pi Dt}}\exp\left[-\left(\frac{z-Wt}{2\sqrt{Dt}}\right)^{2}\right].\label{eq:gaussianSolution}
\end{equation}

Note that, as the subordination transformation acts on time alone,
the same mapping operator $\mathcal{A}$ can be used to map between
spatial moments of the normal and generalised diffusion equations
\begin{eqnarray}
\left\langle \mathbf{r}\right\rangle ^{\left(\mathrm{GDE}\right)}\left(t\right) & = & \mathcal{A}\left\langle \mathbf{r}\right\rangle ^{\left(\mathrm{SDE}\right)}\left(t\right),\\
\left\langle \mathbf{r}\mathbf{r}\right\rangle ^{\left(\mathrm{GDE}\right)}\left(t\right) & = & \mathcal{A}\left\langle \mathbf{r}\mathbf{r}\right\rangle ^{\left(\mathrm{SDE}\right)}\left(t\right),
\end{eqnarray}
where the superscript ``(GDE)'' denotes the generalised diffusion
equation (\ref{eq:generalisedDiffusion}) and ``(SDE)'' denotes
the standard diffusion equation (\ref{eq:normalDiffusion}). Additionally,
the commutation relationship
\begin{equation}
\left[\mathcal{A},\frac{\mathrm{d}}{\mathrm{d}t}\right]\equiv\left[\nu_{\mathrm{loss}}^{\left(\mathrm{free}\right)}+\nu_{\mathrm{trap}}\left(1-\Phi\left(t\right)\ast\right)\right]\mathcal{A},
\end{equation}
also allows the centre of mass (CM) transport coefficients for each
diffusion equation to be related through the subordination transformation
$\mathcal{A}$
\begin{equation}
\mathbf{W}_{\mathrm{CM}}^{\left(\mathrm{GDE}\right)}\left(t\right)=\mathrm{\mathcal{A}}\mathbf{W}_{\mathrm{CM}}^{\left(\mathrm{SDE}\right)}\left(t\right)-\left[\nu_{\mathrm{loss}}^{\left(\mathrm{free}\right)}+\nu_{\mathrm{trap}}\left(1-\Phi\left(t\right)\ast\right)\right]\mathcal{A}\left\langle \mathbf{r}\right\rangle ^{\left(\mathrm{SDE}\right)}\left(t\right),
\end{equation}
\begin{multline}
\mathbf{D}_{\mathrm{CM}}^{\left(\mathrm{GDE}\right)}\left(t\right)=\mathcal{A}\mathbf{D}_{\mathrm{CM}}^{\left(\mathrm{SDE}\right)}\left(t\right)\\
-\frac{1}{2}\left[\nu_{\mathrm{loss}}^{\left(\mathrm{free}\right)}+\nu_{\mathrm{trap}}\left(1-\Phi\left(t\right)\ast\right)\right]\mathcal{A}\left[\left\langle \mathbf{r}\mathbf{r}\right\rangle ^{\left(\mathrm{SDE}\right)}\left(t\right)-\left\langle \mathbf{r}\right\rangle ^{\left(\mathrm{SDE}\right)}\left(t\right)\left\langle \mathbf{r}\right\rangle ^{\left(\mathrm{SDE}\right)}\left(t\right)\right],
\end{multline}
where the CM transport coefficients are defined in terms of spatial
moments according to Eqs. (\ref{eq:CMdrift}) and (\ref{eq:CMdiffusion}).

\section{\label{sec:Conclusion}Conclusion}

We have considered a general phase-space kinetic equation (\ref{eq:boltzmannEquation})
which considers transport of charged particles via both delocalised
and localised states, including collisional trapping, detrapping and
recombination processes. The solution of this model was found analytically
in Fourier-Laplace space which in turn provided analytical expressions
for phase-space averaged spatial and velocity moments. These moments
provided determination of both centre of mass (CM) and flux transport
coefficients. As consequence of the processes of trapping and detrapping,
the free particle CM transport coefficients were found to be transiently
negative for high trapping rates. We have also shown that, in the
hydrodynamic regime, a number of diffusion equations accurately describe
the generalised Boltzmann equation (\ref{eq:boltzmannEquation}).
These include the standard diffusion equation (\ref{eq:normalDiffusionCM}),
the generalised diffusion equation (\ref{eq:generalisedDiffusion})
and, when transport is dispersive, the Caputo fractional diffusion
equation (\ref{eq:fractionalDiffusion}). Finally, we have written
the solution of the generalised diffusion equation (\ref{eq:generalisedDiffusion})
as a subordination transformation (\ref{eq:subordinationTransformation})
from the corresponding solution of a standard diffusion equation (\ref{eq:normalDiffusion}).

The model of focus in this work, Eqs. (\ref{eq:boltzmannEquation})-(\ref{eq:trapLosses}),
was considered only for constant process rates, independent of particle
energy. Extension to higher order balance equations (e.g. momentum
and energy) including energy dependent rates represents the next step
in extending this model. This will facilitate the generalisation of
well known empirical relationships (e.g. Generalised Einstein relations,
Wannier energy relation, mobility expressions) to include combined
localised/delocalised transport systems. Additionally, for our model
to be applied to transport in dense fluids, it is necessary to have
reasonable inputs $\nu_{\mathrm{trap}}$ and $\phi\left(t\right)$.
Although there are many investigations of the trapping, for example
light-particle solvation in the literature \citep{Ceperley1980,Miller2008a,:/content/aip/journal/jcp/99/4/10.1063/1.465198},
including free-energy changes and solvation time-scales, none of these
directly produce an energy-dependent trapping frequency or waiting
time distribution. The ab initio calculation of such capture collision
frequencies and waiting time distributions in liquids and dense gases
remains the focus of our current attention.

\section*{Acknowledgements}

The authors gratefully acknowledge the useful discussions with Prof.
Robert Robson, and the financial support of Australian Research Council
and the Queensland State Government. 

\appendix

\section{\label{sec:R}List of limiting ratios of detrapping and trapping
rates $R$}

\begin{table}[H]
\caption{Specific cases of the dimensionless quantity $R$ that is defined
in Eq. (\ref{eq:Rdefinition}) as the limiting ratio of particle detrapping
and trapping rates. From its alternate definition (\ref{eq:Rrelation}),
it can be seen that $R$ is a function of the difference in recombination
loss rates $\Delta\nu_{\mathrm{loss}}\equiv\nu_{\mathrm{loss}}^{\left(\mathrm{free}\right)}-\nu_{\mathrm{loss}}^{\left(\mathrm{trap}\right)}$.}

\begin{tabular}{|c|c|c|}
\hline 
Trap type & Waiting time distribution $\phi\left(t\right)$ & Limiting ratio of detrapping and trapping rates $R$\tabularnewline
\hline 
\hline 
Instantaneous & $\delta\left(t\right)$ & $1$\tabularnewline
\hline 
Fixed delay & $\delta\left(t-\nu_{\mathrm{detrap}}^{-1}\right)$ & $\frac{\nu_{\mathrm{detrap}}}{\nu_{\mathrm{trap}}}W_{\mathrm{Lambert}}\left[\frac{\nu_{\mathrm{trap}}}{\nu_{\mathrm{detrap}}}\exp\left(\frac{\Delta\nu_{\mathrm{loss}}+\nu_{\mathrm{trap}}}{\nu_{\mathrm{detrap}}}\right)\right]\dagger$\tabularnewline
\hline 
Poisson process & $\nu_{\mathrm{detrap}}\mathrm{e}^{-\nu_{\mathrm{detrap}}t}$ & $\frac{\Delta\nu_{\mathrm{loss}}+\nu_{\mathrm{trap}}-\nu_{\mathrm{detrap}}+\sqrt{\left(\Delta\nu_{\mathrm{loss}}+\nu_{\mathrm{trap}}+\nu_{\mathrm{detrap}}\right)^{2}-4\nu_{\mathrm{detrap}}\Delta\nu_{\mathrm{loss}}}}{2\nu_{\mathrm{trap}}}$\tabularnewline
\hline 
Multiple trapping model & $\alpha\nu_{0}\left(\nu_{0}t\right)^{-\alpha-1}\gamma\left(\alpha+1,\nu_{0}t\right)$ & $\ddagger$\tabularnewline
\hline 
\end{tabular}

$\dagger$ The Lambert W-function is defined as satisfying
\begin{equation}
W_{\mathrm{Lambert}}\left(z\right)\mathrm{e}^{W_{\mathrm{Lambert}}\left(z\right)}\equiv z.
\end{equation}

$\ddagger$ $R$ is the positive solution of the transcendental equation
\begin{equation}
R=-\frac{\alpha\pi}{\sin\alpha\pi}\left(-\frac{\Delta\nu_{\mathrm{loss}}+\nu_{\mathrm{trap}}\left(1-R\right)}{\nu_{0}}\right)^{\alpha}-\alpha\Phi_{\mathrm{Lerch}}\left(\frac{\Delta\nu_{\mathrm{loss}}+\nu_{\mathrm{trap}}\left(1-R\right)}{\nu_{0}},1,-\alpha\right),
\end{equation}
where the Lerch transcendent is defined
\begin{equation}
\Phi_{\mathrm{Lerch}}\left(z,s,a\right)\equiv\sum_{n\geq0}\frac{z^{n}}{\left(n+a\right)^{s}}.\label{eq:lerchDefinition}
\end{equation}
\end{table}

\section{\label{sec:A}List of subordination transformation kernels $A\left(\tau,t\right)$}

\begin{table}[H]
\caption{Specific cases of the integral kernel (\ref{eq:integralKernel}),
$A\left(\tau,t\right)$, used in the subordination transformation
(\ref{eq:subordinationTransformation}) that maps from the solution
of the standard diffusion equation (\ref{eq:normalDiffusion}) to
that of the generalised diffusion equation (\ref{eq:generalisedDiffusion}).}

\begin{tabular}{|c|c|l|}
\hline 
Trap type & Waiting time distribution $\phi\left(t\right)$ & Scaled subordination kernel $\mathrm{e}^{\left(\nu_{\mathrm{loss}}^{\left(\mathrm{free}\right)}+\nu_{\mathrm{trap}}\right)\tau}A\left(\tau,t\right)$\tabularnewline
\hline 
\hline 
Instantaneous & $\delta\left(t\right)$ & $\mathrm{e}^{\nu_{\mathrm{trap}}\tau}\mathrm{e}^{-\nu_{\mathrm{loss}}^{\left(\mathrm{trap}\right)}t}\delta\left(t\right)$\tabularnewline
\hline 
Fixed delay & $\delta\left(t-\nu_{\mathrm{detrap}}^{-1}\right)$ & $\mathrm{e}^{-\nu_{\mathrm{loss}}^{\left(\mathrm{trap}\right)}t}\frac{\left(\nu_{\mathrm{trap}}\tau\right)^{\nu_{\mathrm{detrap}}t}}{\Gamma\left(1+\nu_{\mathrm{detrap}}t\right)}\mathrm{III}_{\nu_{\mathrm{detrap}}^{-1}}\left(t\right)\dagger$\tabularnewline
\hline 
Poisson process & $\nu_{\mathrm{detrap}}\mathrm{e}^{-\nu_{\mathrm{detrap}}t}$ & $\delta\left(t\right)+\frac{1}{t}\mathrm{e}^{-\nu_{\mathrm{detrap}}t}\sqrt{\nu_{\mathrm{trap}}\tau\nu_{\mathrm{detrap}}t}I_{1}\left(2\sqrt{\nu_{\mathrm{trap}}\tau\nu_{\mathrm{detrap}}t}\right)\ddagger$\tabularnewline
\hline 
Multiple trapping model & $\alpha\nu_{0}\left(\nu_{0}t\right)^{-\alpha-1}\gamma\left(\alpha+1,\nu_{0}t\right)$ & $\mathrm{e}^{-\nu_{\mathrm{loss}}^{\left(\mathrm{trap}\right)}t}\left[\frac{1}{t_{\alpha}}l_{\alpha}\left(\frac{t}{t_{\alpha}}\right)\ast g\left(t\right)\right]\mathsection$\tabularnewline
\hline 
\end{tabular}

$\dagger$ The Dirac comb of period $T$ is defined
\begin{equation}
\mathrm{III}_{T}\left(t\right)\equiv\sum_{n\in\mathbb{Z}}\delta\left(t-nT\right).
\end{equation}
$\ddagger$ The modified Bessel function of the first kind of order
$\nu$ is defined
\begin{equation}
I_{\nu}\left(z\right)\equiv\sum_{n\geq0}\frac{\left(\frac{z}{2}\right)^{2n+\nu}}{n!\Gamma\left(1+n+\nu\right)}.
\end{equation}

$\mathsection$ The characteristic time $t_{\alpha}$ is defined
\begin{equation}
\nu_{0}t_{\alpha}\equiv\sqrt[\alpha]{\frac{\alpha\pi}{\sin\alpha\pi}\nu_{\mathrm{trap}}\tau},
\end{equation}
we define in Laplace space the one-sided Lévy density
\begin{equation}
l_{\alpha}\left(p\right)\equiv\mathrm{e}^{-p^{\alpha}},
\end{equation}
and in Laplace space
\begin{equation}
g\left(p\right)\equiv\exp\left[-\alpha\Phi_{\mathrm{Lerch}}\left(-\frac{p}{\nu_{0}},1,-\alpha\right)\right],
\end{equation}
where the Lerch transcendent $\Phi_{\mathrm{Lerch}}\left(z,s,a\right)$
is given by Eq. (\ref{eq:lerchDefinition}).
\end{table}

\bibliography{references,library}

\end{document}